\documentclass[11pt,a4paper]{article}
\pdfoutput=1
\usepackage{jheppub,bm,booktabs,multirow}
\usepackage{xcolor}
\usepackage{overpic}
\usepackage{enumitem} 
\usepackage{verbatim}

\usepackage{adjustbox}
\usepackage{tikz}
\usetikzlibrary{decorations.pathreplacing}
\usetikzlibrary{arrows, arrows.meta}

\allowdisplaybreaks

\makeatletter
\def\@fpheader{~}
\makeatother

\usepackage[Q=yes,pverb-linebreak=no]{examplep}

\newcommand{\slsh}{\!\!\!\slash}

\newcommand{\re}{\text{ref}}

\title{QCD anatomy of photon isolation}
\author[a]{Thomas Becher}
\author[a]{\!, Samuel Favrod}
\author[a]{and Xiaofeng Xu}

\affiliation[a]{Albert Einstein Center for Fundamental Physics, Institut f\"ur Theoretische Physik, Universit\"at Bern,
  Sidlerstrasse 5, CH-3012 Bern, Switzerland}

\emailAdd{becher@itp.unibe.ch}
\emailAdd{favrod@itp.unibe.ch}
\emailAdd{xuxiaofeng@itp.unibe.ch}

\date{\today}

\abstract{To separate the energetic photons produced in hard scattering processes from those from other sources, measurements impose isolation requirements which restrict the hadronic radiation inside a cone around the photon. In this paper, we perform a detailed factorization analysis of the QCD effects associated with photon isolation. We show that for small cone radius $R$, photon isolation effects can be captured by a fragmentation function describing the decay of a parton into a photon accompanied by hadronic radiation. We compute this fragmentation function for different isolation criteria and solve the associated renormalization group equations to resum logarithms of $R$. For small isolation energy, the cone fragmentation function factorizes further, into collinear functions describing energetic quarks and gluons near the cone boundary and functions encoding their soft radiation emitted into the cone. Based on this factorization we also resum the non-global logarithms of the ratio of the photon energy and the isolation energy, so that we control all logarithmically enhanced terms in the cross section. In this limit, we provide a simple formula to convert NNLO cross section results from smooth-cone isolation to fixed-cone isolation.}

\begin{document}

\maketitle

\section{Introduction}

An important category of physics probes at high-energy colliders are processes with electroweak bosons in the final state. Among these, photons present special challenges: since they are massless, they are abundant and are produced not only during the hard interaction, but can also arise as secondary emissions during jet fragmentation, hadronization and hadron decay. The fragmentation process involves non-perturbative physics encoded in photon fragmentation functions, originating from partons becoming collinear to the photon.  

To reduce the contribution from secondary emissions, experiments impose isolation requirements. To isolate a hard photon they put a cone of angular size $R$ around it and restrict the hadronic energy inside the cone to be lower than a certain cutoff $E_0$. How this cutoff is imposed depends on the isolation criterion. The simplest way is to impose a constraint on the total hadronic energy $E_{\rm tot}(R)$ inside the cone. At an $e^+e^-$ collider one requires
\begin{equation}\label{eq:fixed}
  \text{fixed-cone isolation:}\quad   E_{\rm tot}^{\rm cone}(R) < E_0= \epsilon_\gamma E_\gamma\, 
\end{equation}
and the quantity $R$ corresponds to the opening half-angle of the cone, i.e.\ a particle is inside the cone if $\theta < R$, where $\theta$ is the angle between the particle and the photon. At hadron colliders, one instead imposes the constraint on the total transverse energy $E_T$ inside the cone and defines $E_0= \epsilon_\gamma E^\gamma_T$.\footnote{The transverse energy of a particle is defined as $E_T = E \sin \theta_b$, where $\theta_b$ is its angle with respect to the beam axis.} In the following, we will use the term energy to refer to either the conventional or the transverse energy, depending on the collider under consideration. At a hadron collider a particle is inside the cone if $r<R$ with $r=\sqrt{(\Delta \eta)^2+(\Delta \phi)^2}$. Here, $\Delta \eta$ and $\Delta \phi$ are the pseudorapidity and azimuthal angle differences between the photon and the particle. Fixed-cone isolation is used in all experimental measurements by ATLAS~\cite{ATLAS:2017nah,ATLAS:2019buk,ATLAS:2019iaa} and CMS~\cite{CMS:2014mvm,CMS:2018qao,CMS:2019jlq}, but with this isolation criterion the cross section computations need to include the non-perturbative photon fragmentation functions to be collinear finite. The photon fragmentation functions are poorly known and the presence of final-state collinear divergences complicates the perturbative calculations.

Frixione \cite{Frixione:1998jh} has introduced an alternative isolation criterion designed to eliminate radiation collinear to the photon. Rather than restricting the radiation inside a fixed cone of radius $R$, it imposes
\begin{equation}\label{eq:smooth}
  \text{smooth-cone isolation:}\quad   E_{\rm tot}^{\rm cone}(r) <  E_0(r)=\epsilon_\gamma E_\gamma \left(\frac{1-\cos r}{1-\cos R}\right)^n \,
\end{equation}
for all $r<R$ where the parameter $n$ must be chosen to be $n\geq \frac{1}{2}$. As is obvious from this definition, the isolation becomes stricter as particles get more collinear to the photon and together with collinear radiation, the smooth-cone isolation also eliminates the non-perturbative fragmentation function. Having infrared finite cross sections without the need to subtract collinear final state singularities is a significant technical simplification and in the past all next-to-next-to-leading order (NNLO) computations of photon production were carried out imposing smooth-cone isolation. Such computations are by now available not only for inclusive-photon~\cite{Campbell:2016lzl,Chen:2019zmr}, photon-plus-jet~\cite{Campbell:2017dqk,Chen:2019zmr}
 di-photon~\cite{Catani:2011qz,Campbell:2016yrh,Catani:2018krb,Gehrmann:2020oec,Chawdhry:2021hkp,Badger:2021ohm}
 and even tri-photon  production~\cite{Chawdhry:2019bji,Kallweit:2020gcp}. However, the finite granularity of the detectors makes it impossible to directly implement the criterion \eqref{eq:smooth} experimentally. While a discretized version was studied for the LHC \cite{SM:2010nsa} and, following earlier work on democratic clustering \cite{Glover:1993xc}, new isolation criteria based on jet substructure \cite{Hall:2018jub} were proposed, all LHC measurements currently impose fixed-cone isolation.
 
 To compare to the experimental results, the above theoretical papers choose parameters $n$ and $\epsilon_\gamma$ of the smooth-cone isolation to mimic the fixed-cone isolation applied in the measurement. In the literature, a variety of parameter choices is found, typically motivated by next-to-leading order (NLO) computations, which are available both with fixed cone and smooth-cone isolation \cite{Catani:2002ny}. Given that the two isolation criteria are qualitatively different, the situation is unsatisfactory, especially since the experimental measurements now reach few per-cent accuracy. The paper \cite{Gehrmann:2020oec} has shown that photon isolation is a substantial source of uncertainty in precision calculations and has advocated the use of a hybrid isolation scheme,  in which a small smooth cone is placed into the center of a fixed cone, to mitigate this problem \cite{Siegert:2016bre,Gehrmann:2020oec}. Very recently the antenna subtraction method \cite{Gehrmann-DeRidder:2005btv,Daleo:2006xa,Currie:2013vh} has been extended to final state singularities \cite{Gehrmann:2022cih} and by now the first NNLO fixed-cone result is available \cite{Chen:2022gpk}, eliminating the mismatch between the prediction and the experimental measurements. 
  
 By construction, the isolation requirement introduces low scales into the cross section, which leads to logarithmically enhanced higher-order terms which can spoil perturbative predictions. The paper \cite{Catani:2002ny} has computed the photon cross section at NLO and has shown that for small radius the prediction for the isolated cross section becomes larger than the inclusive cross section, clearly indicating a breakdown of fixed-order perturbation theory. The leading $\ln(R)$ terms were then resummed in \cite{Catani:2013oma} curing this pathology. The resummation can be obtained by evolving the fragmentation contribution from the hard scale $\mu_h \sim E_\gamma$ to the scale $\mu_j \sim R\,E_\gamma$ associated with the invariant mass of the radiation in and around the cone.  In our paper we perform a detailed factorization analysis of the QCD effects arising due to photon isolation in the framework of Soft-Collinear Effective Theory (SCET) \cite{Bauer:2000yr,Bauer:2001yt,Beneke:2002ph}. In addition to the logarithms of the isolation-cone radius $R$, we will also resum logarithms of $\epsilon_\gamma = E_0/E_\gamma$. These involve the scale $\mu_0 \sim R\,E_0$, which is typically quite low for experimentally imposed isolation criteria. 
 
 The factorization of the photon cross section involves two steps. First, we show that for small isolation cone size $R$, the isolation effects can be captured by an isolation fragmentation function, i.e. a fragmentation function which describes the fragmentation of a parton into a photon plus the accompanying collinear radiation constrained by the isolation criterion. The fragmentation function factorization makes it easy to study the effect of different isolation criteria and the dependence on isolation parameters. It  also makes it possible to convert results obtained in one isolation scheme to another, since the cross section difference is driven by the difference in the associated fragmentation functions. Generalized fragmentation functions similar to the one we introduce have been used in a variety of other contexts starting with \cite{Procura:2009vm}, who considered hadron fragmentation inside a jet.  The fragmentation function approach can be used to resum logarithms of the isolation-cone radius $R$ by solving evolution equations to evolve from the hard scale $\mu_h\sim E_\gamma$ down to the typical scale of the fragmentation function $\mu_j\sim E_\gamma R$. The same technique has been used earlier for inclusive jet production, where one can consider fragmentation into a jet to resum logarithms of $R$, as was done in SCET in \cite{Kang:2016mcy,Dai:2016hzf,Cal:2019hjc}, following analogous computations in QCD factorization \cite{Dasgupta:2014yra,Dasgupta:2016bnd}. The resummation of $\ln(R)$ terms has also been studied for exclusive jet production \cite{Chien:2014nsa,Becher:2015hka,Becher:2016mmh} and inclusive jet production near threshold \cite{Liu:2017pbb,Liu:2018ktv}, where the $R$ dependence is captured by jet functions instead of fragmentation functions.
 
 In addition to the collinear scale $\mu_j\sim E_\gamma R$, the fragmentation function involves the scale $\mu_0\sim E_0 R$ associated with radiation inside the isolation cone. We will show that in the limit of small $E_0$, the cone fragmentation function itself factorizes. This second factorization step then allows us to also resum the logarithms of $\epsilon_\gamma$, as  shown in \cite{Balsiger:2018ezi}. The isolation cone is obviously a non-global observable \cite{Dasgupta:2001sh} and we use the RG approach of \cite{Becher:2015hka,Becher:2016mmh} implemented in the code {\sc NGLresum} \cite{Balsiger:2020ogy} to resum them. In this way we control all large logarithms associated with photon isolation.
  
 Our paper is organized as follows. We first present factorization a theorem for photon production with a small isolation cone in Section \ref{sec:SmallR}. The leading-order fragmentation functions which encode the isolation are computed in Section \ref{sec:fragmentation}. Using these, we can study the differences among isolation criteria and their parameter dependence. We can also resum contributions enhanced by logarithms of the cone radius $R$ by solving the Dokshitzer-Gribov-Lipatov-Altarelli-Parisi (DGLAP) evolution equation for the fragmentation functions as explained in Section \ref{sec:lnRres}. We then discuss the factorization of the fragmentation function in the limit of small isolation energy in Section \ref{sec:smallIsolationEnergy}. The leading jet function arising in this factorization theorem is computed in Section \ref{sec:jet}, together with the function describing the soft radiation into the cone. These results are then used to derive a formula to convert smooth-cone cross section results to fixed-cone isolation in the limit of small $\epsilon_\gamma$. In Section \ref{sec:resummation} we then perform the resummation of logarithms of $\epsilon_\gamma$. We summarize our results and conclude in Section \ref{sec:conclusion}.

\begin{figure}[t!]
\centering
\includegraphics[width=0.5\linewidth]{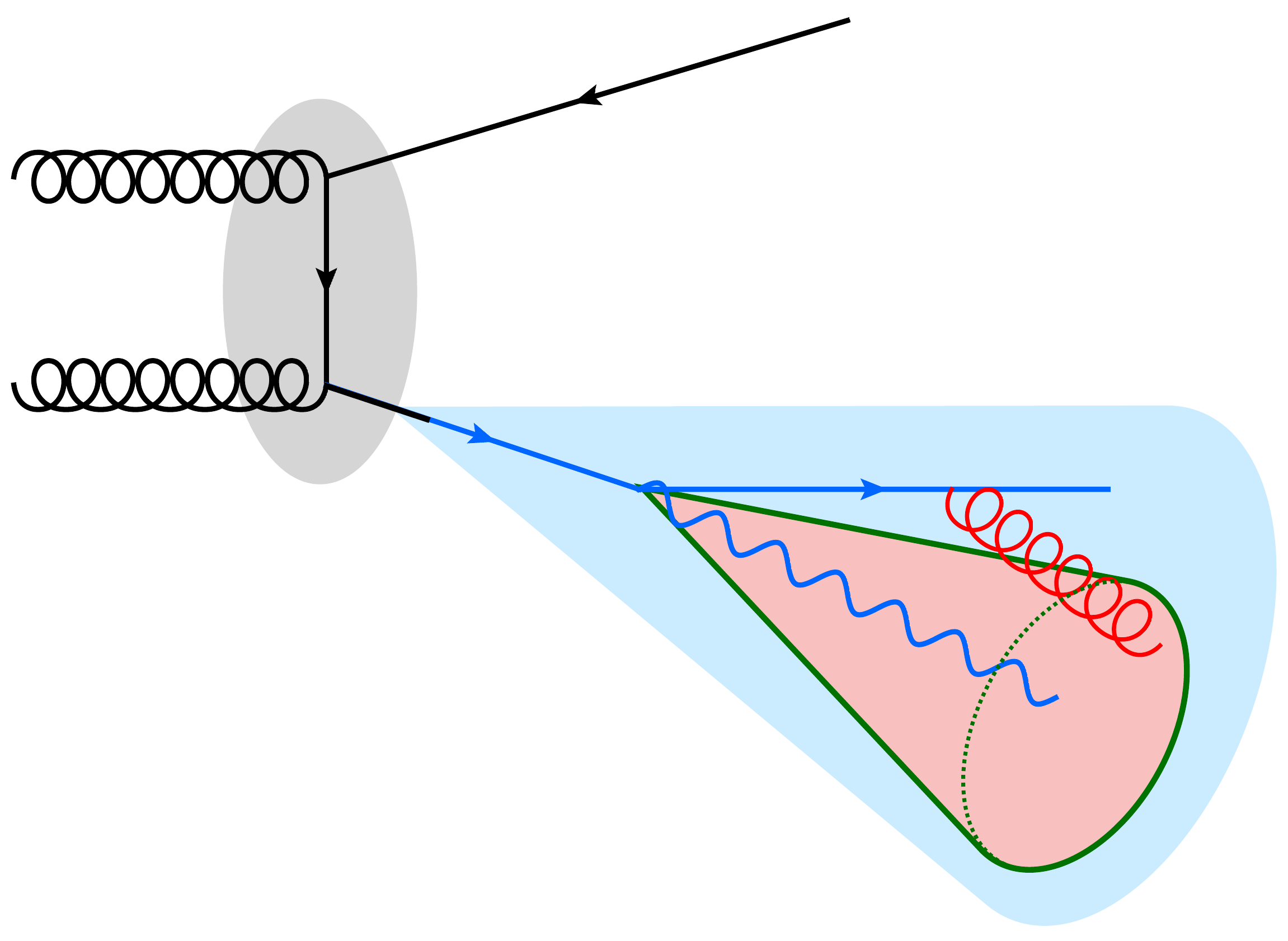}
\caption{Pictorial representation of the factorization theorems \eqref{factorizationFormulaSmallCone} and \eqref{eq:deffrag}. The gray blob represents the hard function, which describes the production of an energetic parton, which then fragments into a photon plus additional radiation (blue region), as encoded by the cone fragmentation function $\mathcal{F}_{i\to \gamma}$ in \eqref{factorizationFormulaSmallCone}. For small isolation energy, this function factorizes further. The energetic partons (blue lines) produced in the fragmentation are part of the jet function $\bm{\mathcal{J}}_{\!\! i \to \gamma+l}$ and must lie outside the isolation cone. These partons can then radiate soft partons (red) into the isolation cone (green). This radiation is encoded in the functions $\bm{\mathcal{U}}_{l}$, which depend on the directions and color charges of the energetic partons.}
\label{fig:fact}
\end{figure}

\section{Factorization for isolated photon production at small cone radius $R$}\label{sec:SmallR}

For small isolation cone radius $R$ a factorization theorem for isolated photon production was presented in \cite{Balsiger:2018ezi}. It reads
\begin{align}\label{factorizationFormulaSmallCone}
\frac{\text{d}\sigma (E_0,R)}{\text{d}E_\gamma}&= \frac{\text{d}\sigma^{\rm dir}_{\gamma+X}}{\text{d}E_\gamma}\nonumber\\
&\hspace{1cm}+\sum_{i=q,\bar{q},g} \int dz \frac{\text{d}\sigma_{i+X}}{\text{d}E_i} \mathcal{F}_{i\to \gamma}(z, E_\gamma, E_0, R) +\mathcal{O}(R)\,,
\end{align} 
where the isolation-cone fragmentation function $\mathcal{F}_{i\to \gamma}$ describes the fragmentation of the hard parton with energy $E_i$ into a photon with energy $E_{\gamma} = z E_i$ plus accompanying hadronic radiation which is restricted to have energy smaller than $E_0$ inside the cone, see Figure \ref{fig:fact}. The precise definition of this function is given below. The quantity $\sigma^{\rm dir}_{\gamma+X}$ is the perturbative cross section for producing a photon without imposing any isolation. The direct part is not collinear safe by itself, but its divergences cancel against the fragmentation part of the cross section. A more compact (and slightly more general) way of writing formula \eqref{factorizationFormulaSmallCone} is
\begin{align}\label{factorizationFormulaSmallConeAlt}
\frac{\text{d}\sigma (\epsilon_\gamma,R)}{\text{d}E_\gamma}& =  \sum_{i=\gamma,q,\bar{q},g} \int dz \frac{\text{d}\sigma_{i+X}}{\text{d}E_i} \mathcal{F}_{i\to \gamma}(z, E_\gamma, E_0, R)\,.
\end{align}
Note that in this second form the sum over partons includes the photon. Throughout our paper, we work at leading order in the electromagnetic coupling $\alpha$ and neglect its running so that we have
\begin{equation}
\mathcal{F}_{\gamma\to \gamma} = \delta(1-z)\,,
\label{eq:gammafrag}
\end{equation}
which leads back to the original form \eqref{factorizationFormulaSmallCone} of the equation. 

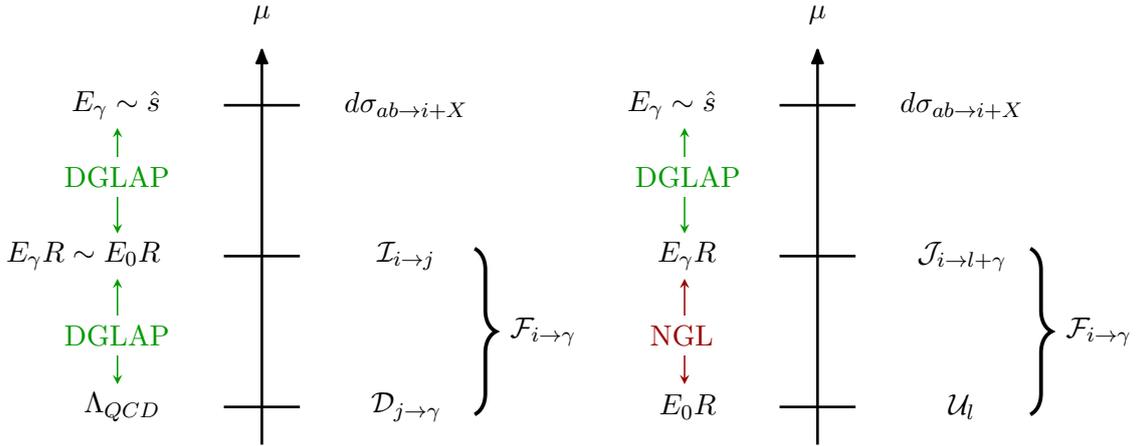
\begin{figure}[t!]
\begin{adjustbox}{center}

\begin{tabular}{cc}

\begin{tikzpicture}


\draw[-{Latex[round]},line width=0.35mm](0,-2.5)--(0,2.75);
\node at (0,3.2) {$\mu$};
\draw[black,line width=0.35mm] (-0.5,-2)--(0.5,-2);
\draw[black,line width=0.35mm] (-0.5,0)--(0.5,0);
\draw[black,line width=0.35mm] (-0.5,2)--(0.5,2);

\node at (1.9,2) {$d \sigma_{ab \to i + X} $};
\node at (1.9,0.0) {$\mathcal{I}_{i \to j} $};
\node at (1.9,-2){$\mathcal{D}_{j \to \gamma} $};

\draw [decorate,decoration={brace,amplitude=8pt},line width=0.5mm] (2.8,0.1) --  (2.8,-2.1);
\node at (3.7,-1) {$\mathcal{F}_{i\to \gamma}$};

\node[anchor=east] at (-1.2,2.0) {$E_{\gamma} \sim \hat{s} $};
\node[anchor=east] at (-1.2,0.0) {$E_{\gamma} R \sim E_0 R $};
\node[anchor=east] at (-1.2,-2.) {$ \Lambda_{QCD} $};


\draw[>=stealth, <->,green!60!black,line width=0.2mm] (-1.9,1.7) -- (-1.9,0.3);

\node[text width=2cm,green!60!black,fill=white] at (-1.6,1.05) {DGLAP};


\draw[>=stealth, <->,green!60!black,line width=0.2mm] (-1.9,-1.7) -- (-1.9,-0.3);

\node[text width=2cm,green!60!black,fill=white] at (-1.6,-1.05) {DGLAP};

\end{tikzpicture} &
\begin{tikzpicture} 


\draw[-{Latex[round]},line width=0.35mm](0,-2.5)--(0,2.75);
\node at (0,3.2) {$\mu$};
\draw[black,line width=0.35mm] (-0.5,-2)--(0.5,-2);
\draw[black,line width=0.35mm] (-0.5,0)--(0.5,0);
\draw[black,line width=0.35mm] (-0.5,2)--(0.5,2);

\node at (1.9,2) {$d \sigma_{ab \to i + X} $};
\node at (1.9,0.0) {$\mathcal{J}_{i \to l + \gamma} $};
\node at (1.9,-2) {$\mathcal{U}_{l} $};

\draw [decorate,decoration={brace,amplitude=8pt},line width=0.5mm] (2.8,0.1) --  (2.8,-2.1);
\node at (3.7,-1) {$\mathcal{F}_{i\to \gamma}$};

\node[anchor=east] at (-1.2,2) {$E_{\gamma} \sim \hat{s} $};
\node[anchor=east] at (-1.2,0.0) {$E_{\gamma} R  $};
\node[anchor=east] at (-1.2,-2) {$E_0 R $};


\draw[>=stealth, <->,green!60!black,line width=0.2mm] (-1.75,1.7) -- (-1.75,0.3);

\node[text width=2cm,green!60!black,fill=white] at (-1.4,1.05) {DGLAP};


\draw[>=stealth, <->,red!60!black,line width=0.2mm] (-1.75,-1.7) -- (-1.75,-0.3);

\node[text width=2cm,red!60!black,fill=white] at (-1.2,-1.05) {NGL};
\end{tikzpicture}

\end{tabular}

\end{adjustbox}
\caption{The scales arising in the factorization theorems \eqref{factorizationFormulaSmallCone} and \eqref{eq:deffrag}, together with the type of RG evolution needed to resum the associated logarithms. On the left we show the factorization when $R$ is small and the isolation energy $E_0$ is parametrically of the same size as the photon energy $E_\gamma$. On the right we show the factorization for $R\ll 1$ and $\epsilon_\gamma = E_0/E_\gamma \ll 1$. In this limit non-perturbative fragmentation effects are suppressed by $\epsilon_\gamma$.}
\label{fig:scales}
\end{figure}

The fact that the photon cross section involves a fragmentation function which describes the conversion of a parton into a photon plus collinear partons is well known \cite{Koller:1978kq,Laermann:1982jr,Owens:1986mp}, see \cite{Kaufmann:2017lsd} for a recent review. What is different in our case is the definition and role of the fragmentation function. The standard fragmentation functions encode non-perturbative effects in photon production, while our function includes all physics associated with photon-isolation and therefore also has a perturbative component. The function $\mathcal{F}_{i\to \gamma}$ in \eqref{factorizationFormulaSmallConeAlt} describes the fragmentation of the energetic parton $i$ into a photon in the presence of the isolation cone, up to corrections suppressed by powers of $R$. Since we expand in small $R$, the isolation cone radius is set to zero when the partonic cross section $d\sigma_{i+X}$ is computed, which leads to infrared (IR) divergences which match the UV divergences of the fragmentation function. 

In SCET, the fragmentation function is obtained as a matrix element of collinear fields, whose light-cone momentum components scale as
\begin{equation}\label{pertColl}
( n\cdot p, \bar{n}\cdot p, p_\perp ) \sim E_\gamma ( R^2, 1, R)  \,,
\end{equation}
where $n^\mu$ is a reference vector along the direction of the fragmenting parton. The $\perp$-directions are perpendicular to the fragmenting parton, not the beam.  In this section, we do not consider the hierarchy between $E_\gamma$ and $E_0$, i.e. we treat $E_\gamma \sim E_0$, corresponding to the situation shown on the left-hand side of Figure \ref{fig:scales}. The limit of small $E_0$ will be considered later in Section \ref{sec:smallIsolationEnergy} and will lead to an additional factorization of the fragmentation functions.

The definition of the fragmentation functions for quarks and gluons reads
\begin{align}\label{eq:fragq}
\frac{{n\!\!\!/}_{\alpha\beta}}{2}\delta^{ab} \mathcal{F}_{q\to \gamma}&(z, E_\gamma, E_0, R,\mu) \nonumber\\  &=\sum\hspace{-0.67cm}\int\displaylimits_{\gamma+X}  \langle 0 | \chi^a_{q\alpha}(0) |\gamma\!+\! X \rangle \langle \gamma\!+\! X |\,\bar{\chi}^b_{q\beta}(0)\,|0\rangle \, \theta(2E_0 - \bar{n}\cdot p_{X}^{\rm in}) \,\delta\!\left(z  - \frac{\bar{n}\cdot p_\gamma}{Q}\right) \nonumber \\
 & \hspace{2.5cm}\,(2\pi)^{d-1}\delta\!\left(Q-\bar{n}\cdot(p_\gamma+p_X)\right)\, \delta^{(d-2)}(p^\perp_\gamma+p^\perp_X)   \,, \\
\label{eq:fragg}
 -g^\perp_{\alpha\beta}\, g_s^2 \delta^{ab} \mathcal{F}_{g\to \gamma}&(z, E_\gamma, E_0, R,\mu) \nonumber\\ &=\sum\hspace{-0.67cm}\int\displaylimits_{\gamma+X} \langle 0 | \mathcal{A}^{\perp a}_\alpha(0) |\gamma\!+\! X \rangle \langle \gamma\!+\! X |\,\mathcal{A}^{\perp b}_\beta(0)\,|0\rangle \, \theta(2E_0 - \bar{n}\cdot p_{X}^{\rm in})\,\delta\!\left(z - \frac{\bar{n}\cdot p_\gamma}{Q}\right) \nonumber \\
 &\hspace{2.5cm}\,(2\pi)^{d-1} Q\, \delta\!\left(Q-\bar{n}\cdot(p_\gamma+p_X)\right) \, \delta^{(d-2)}(p^\perp_\gamma+p^\perp_X) \,,
  \end{align}
where we sum over states $X$ containing collinear QCD partons and integrate over the phase space of the partons in $X$ and the photon. The fields $\chi_{q} = W^\dagger_c \psi_{c,q}$ and $ \mathcal{A}^{\perp a}_\alpha t^a = W^\dagger_c i D_{c,\alpha}^\perp W_c$ are the collinear quark and gluon fields in SCET times their associated collinear Wilson lines $W_c$. The indices $a, b$ and $\alpha,\beta$ are associated with color and spin, respectively. The coupling $g_s$ is the bare strong coupling constant. The Wilson lines $W_c$ make the fields $\chi_{q}$ and $\mathcal{A}_\mu^\perp$ invariant under collinear gauge transformations and are a product of a QED and a QCD Wilson line. The total momentum  of the partons inside the cone of radius $R$ is denoted by  $p_{X}^{\rm in}$ and its large component, not the energy, is bounded by the isolation criterion. Up to power corrections in $R$ the large component $\bar{n}\cdot p$ is equal to twice the energy. The large light-cone component of the momentum of the incoming parton is $Q$ and is given in terms of photon energy as $Q = 2E_\gamma/z$. We have written the constraint for fixed-cone isolation \eqref{eq:fixed}, but it can easily be adapted for smooth-cone case \eqref{eq:smooth}.

\begin{figure}[t!]
\centering
\includegraphics[width=0.7\linewidth]{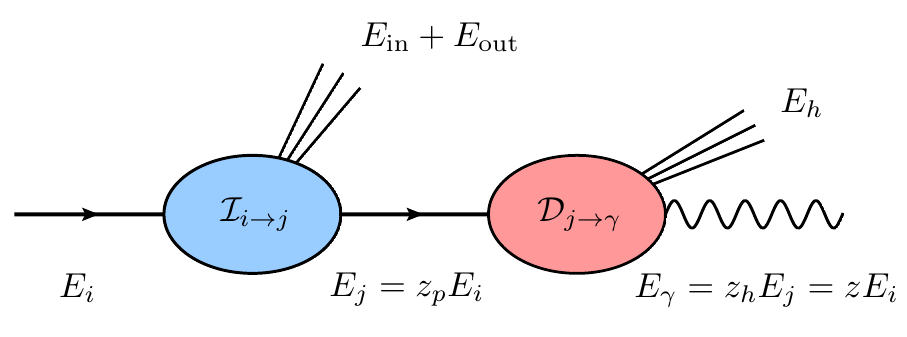}
\caption{Kinematics of the factorization \eqref{eq:nonpertFrag} of the cone fragmentation function $\mathcal{F}_{i\to \gamma}$ into a perturbative and non-perturbative part. The radiated partons in the perturbative part $\mathcal{I}_{i\to j}$ can be inside or outside the isolation cone, while the non-perturbative radiation in $\mathcal{D}_{j \to \gamma}$ is always inside. The perturbative momentum fraction is $z_p = z /z_h$.}
\label{fig:fragfact}
\end{figure}

In general, the fragmentation functions  \eqref{eq:fragq} and \eqref{eq:fragg} also contain a non-perturbative component from partons whose momentum scales as 
\begin{equation}\label{nonpert}
( n\cdot p, \bar{n}\cdot p, p_\perp ) \sim E_\gamma ( \lambda^2, 1, \lambda)  \,,
\end{equation}
with $\lambda \sim \Lambda_{\rm QCD}/E_\gamma$.
These energetic partons are highly collinear to the photon. They are therefore always inside the isolation cone and their energy is constrained. After integrating out the perturbative modes \eqref{pertColl}, one ends up with a low-energy effective theory containing only the modes \eqref{nonpert} and the fragmentation function becomes of convolution of perturbative coefficients $\mathcal{I}_{i\to j}$ times non-perturbative fragmentation functions $\mathcal{D}_{j \to \gamma}$. The associated two-step fragmentation process is depicted in Figure \ref{fig:fragfact}. For fixed-cone isolation, the associated factorization formula reads 
 \begin{multline}\label{eq:nonpertFrag}
    \mathcal{F}_{i\to \gamma}(z, E_\gamma, E_0, R,\mu) = \sum_{j=\gamma,q,\bar{q},g} \int_{z}^1\! \frac{dz_h}{z_h} \int\! dE_{\rm in}\, 
    \theta\!\left(E_0 - E_{\rm in} -\frac{1-z_h}{z_h} E_\gamma\right)\\
    \mathcal{I}_{i\to j}(z/z_h, E_\gamma, E_{\rm in}, R, \mu)\, \mathcal{D}_{j \to \gamma}(z_h,\mu)\,.
\end{multline}
The $\theta$-function is due to photon isolation and constrains the energy inside the cone, which gets contributions from perturbative partons in $\mathcal{I}_{i\to j}$ as well as the non-perturbative partons in $\mathcal{D}_{j \to \gamma}$, which carry the hadronic energy
\begin{equation}
    E_h = (1-z_h) E_j = (1-z_h) \cdot z_p \cdot E_i =  \frac{1-z_h}{z_h} E_\gamma\,,
\end{equation}
where we used that $z_p = z/z_h$, see Figure \ref{fig:fragfact}. While the non-perturbative partons are always inside the cone, the perturbative ones scaling as \eqref{pertColl} can be inside or outside. The constraint only acts on the inside part $E_{\rm in}$ of their energy.
The constraint on the energy inside the cone implies that $z_h>1/(1+\epsilon_\gamma)$. In the limit $\epsilon_\gamma\to 0$, the $z_h$ integration in \eqref{eq:nonpertFrag} no longer has any support.
In this situation, the only contribution arises from $j=\gamma$
\begin{equation}\label{photontophoton}
    \mathcal{D}_{\gamma \to \gamma}(z_h,\mu) = \delta(1-z_h) \,,
\end{equation}
rendering the fragmentation purely perturbative up to  power corrections in $\epsilon_\gamma$. The limit of small isolation energy will be considered in detail below.

The scaling \eqref{nonpert} and the structure of \eqref{eq:nonpertFrag} make it clear why there is no non-perturbative contribution for smooth-cone isolation \eqref{eq:smooth}. Since the non-perturbative partons \eqref{nonpert} are very close to the center of the isolation cone, they are not allowed to carry any energy since $E_0(r) \to 0$ for $r\to 0$. This enforces $z_h \to 1$ and the integral over $z_h$ has no  support and the only contribution arises again from $\mathcal{D}_{\gamma \to \gamma}$ in \eqref{photontophoton}. The smooth-cone fragmentation function is purely perturbative up to corrections suppressed by $\Lambda_{\rm QCD}/E_\gamma$.

The factorization formula \eqref{factorizationFormulaSmallCone} is only valid up to corrections suppressed by the cone radius $R$, but has the advantage that it captures all dependence on photon isolation. As such it is well suited to analyze the dependence of cross sections on isolation parameters and can also be used to convert a result from one isolation criterion to another. We may, for example, convert a result computed using Frixione isolation to a result in fixed-cone isolation by evaluating the difference of the relevant fragmentation functions. A second advantage of the factorization \eqref{factorizationFormulaSmallCone} is that it separates the hard scale $E_\gamma\sim \sqrt{\hat{s}}$ from the collinear scale $E_\gamma R$ associated with the fragmentation. This enables us to use renormalization-group (RG) evolution to resum logarithms of $R$, the ratio of the two scales, as discussed in detail in the next sections.

\section{Isolation fragmentation functions}\label{sec:fragmentation}

We will now analyze the factorization discussed in Section \ref{sec:SmallR} in more detail. Let us start by evaluating the isolation fragmentation functions at $\mathcal{O}(\alpha_s)$. At this order, the only nontrivial fragmentation process is $q \to \gamma(k)+ q(p) $ and the matrix element in \eqref{eq:fragq} is the usual splitting function in $d=4-2 \epsilon$ dimensions
\begin{align}\label{eq:splitting}
 \langle 0 | \chi^b_{\beta}(0)| \gamma \! +\!q \rangle \langle \gamma\!+\!q | \bar{\chi}^a_{\alpha}(0)| 0 \rangle &=\delta^{ab} \left (\frac{ n \slsh }{2}\right)_{\alpha \beta} \frac{e^2 Q_q^2\left( (d-2) (  \bar{n}\cdot k)^2+ 4\, \bar{n} \cdot k \,\bar{n}\cdot p  +4(\bar{n} \cdot  p  )^2 \right)}{2 p\cdot k \ \bar{n} \cdot k }\,,\nonumber\\
 & = \delta^{ab} \left (\frac{ n \slsh }{2}\right)_{\alpha \beta} \frac{e^2 Q_q^2\, Q}{k\cdot p} \left[ \frac{1+(1-z)^2}{z} -\epsilon\, z\right]\,,
 \end{align}
where we have denoted the photon momentum by $k$ and the quark momentum by $p$ and the charge $Q_q$ is $+2/3$ for up-type and $-1/3$ for down-type quarks. In the second line the large light-cone components were written as $k\cdot \bar{n}= z\,Q$ and $p\cdot \bar{n}= (1-z)\,Q$. The expression in square brackets is the spin averaged splitting kernel in $d$ dimensions. To obtain the fragmentation function, we need to integrate the matrix element \eqref{eq:splitting} over the phase space of the photon and quark in the presence of the kinematic constraints in \eqref{eq:fragq}. Expanding away components which are power suppressed according to \eqref{pertColl} , the cone constraint is formulated in terms of the angular quantity
\begin{equation}
\delta_{\gamma q}^2 = \frac{2p\cdot k}{\bar{n}\cdot p\, \bar{n}\cdot k} \,,
\end{equation}
which scales as $\mathcal{O}(R^2)$. Up to higher order terms, we can approximate
\begin{equation}
\delta_{\gamma q} \approx \tan\Big(\frac{\theta_{\gamma q}}{2}\Big) \approx \frac{\theta_{\gamma q}}{2}\, .
\end{equation}
For the fragmentation process $q \to  \gamma(k) + q(p) $, the isolation cone constraint in \eqref{eq:fragq} takes the explicit form
\begin{equation}\label{eq:constraintq}
\theta(2E_0 - \bar{n}\cdot p_{X}^{\rm in}) = \theta(\delta^2- \delta_{\gamma q}^2)\, \theta(2E_0 - \bar{n}\cdot p)+ \theta(\delta_{\gamma q}^2- \delta^2)\,.
\end{equation}
The first term on the right-hand side imposes an energy constraint if the quark is inside the cone. The relation of $\delta$ to the cone size $R$ depends on the collider. In the limit of small $R$ we have
\begin{equation}
\begin{aligned}
\text{$e^+ e^-$ collider: }& & \delta &= R\,,\\
\text{proton collider: }& & \delta &= R \sin \theta_\gamma = R/\cosh(\eta_\gamma)\,.
\end{aligned}
\end{equation}
The hadron-collider result follows from analyzing $r<R$ with $r^2=(\Delta \eta)^2+(\Delta \phi)^2$ near the limit where the quark is collinear to the photon. To present results independent of the collider, we will express them in terms of the quantities $\delta$ and $Q$. For the product of the two at a hadron collider, we have
\begin{equation}
Q \delta = \frac{2 E_\gamma^T}{z \sin \theta_\gamma } R \sin \theta_\gamma = \frac{2 E_\gamma^T}{z} R\,,
\end{equation}
while we get $Q \delta = 2E_\gamma R/z$ at a lepton collider. 

Due to \eqref{eq:constraintq} the leading-order fragmentation function can naturally be split into two terms, depending on whether the quark in the final state is inside or outside the isolation cone
\begin{equation}
    \mathcal{F}_{q\to \gamma}(z,E_\gamma, E_0, R,\mu) =  \mathcal{F}^{\rm in}_{q\to \gamma}(z, E_\gamma, E_0, R ,\mu) +  \mathcal{F}^{\rm out}_{q\to \gamma}(z,R\, E_\gamma,\mu)\,,
\end{equation}
where the outside part is independent of the isolation. The bare result for the outside part reads
\begin{equation}\label{eq:Fout}
 \mathcal{F}^{\rm out}_{q\to \gamma}(z,R\, E_\gamma)=\frac{\alpha_{\text{EM}}\, Q_q^2 }{2 \pi} \left\{ P(z) \left[ \frac{1}{\epsilon}   - \ln\!\left( \frac{\delta^2 Q^2}{\mu^2} (1-z)^2 z^2\right)\right] -z \right\}\,,
\end{equation}
with the $d=4$ splitting kernel
\begin{align}\label{eq:splittingkernel}
P(z)= \frac{1+(1-z)^2}{z} \, 
\end{align}
and after expressing the bare electromagnetic coupling $\alpha_0$ through the $\overline{\rm MS}$ result via $\alpha_0 = Z_\alpha \alpha_{\rm EM} (\mu^2\,e^{\gamma_E}/(4\pi))^{\epsilon}$, where $\gamma_E$ is the Euler-Mascheroni constant. The renormalized result is then obtained by subtracting the divergence in \eqref{eq:Fout}.
The inside part for smooth-cone isolation \eqref{eq:smooth} is given by
\begin{equation}\label{eq:FinSmooth}
 \mathcal{F}^{\rm in}_{q\to \gamma}(z, E_\gamma, E_0, R ,\mu)  =\frac{\alpha_{\text{EM}} Q_q^2}{2 \pi} P(z) \frac{1}{n} \ln\!\left( \frac{ z\, \epsilon_ \gamma }{1-z}\right) \theta\!\left(z-\frac{1}{1+\epsilon_\gamma} \right),
\end{equation}
where $n$ is the exponent parameter of the smooth-cone isolation condition \eqref{eq:smooth}. 
The function $ \mathcal{F}^{\rm in}_{q\to \gamma}$ is finite and independent of of the cone radius, while the outside part has logarithmic $R$ dependence tied to its divergence. As it should be, the total fragmentation function has a divergence proportional to the splitting kernel. Due to the constraint on the inside energy, the inside fragmentation function has only support for large enough $z$ and vanishes in the limit $\epsilon_\gamma\to 0$. Expanding around this limit, we find
\begin{equation}\label{eq:FinSmoothEx}
 \mathcal{F}^{\rm in}_{q\to \gamma}(z, E_\gamma, E_0, R ,\mu)  =\frac{\alpha_{\text{EM}} Q_q^2}{2 \pi} \frac{1}{n}\, \epsilon_\gamma\, \delta(1-z) + \mathcal{O}( \epsilon_\gamma^2) \,.
\end{equation}
The $\epsilon_\gamma$ suppression is expected since the collinear quark becomes soft and soft quarks are power suppressed. 

\begin{table}
\centering
\begin{tabular}{ccc}\hline & & \\
$\sqrt{s} = 13\,{\rm TeV}$ & $E_T^\gamma > E_T^{\rm min}= 125\,{\rm GeV}$ & $|\eta_\gamma| < 2.37$\\[10pt]
 NNPDF23\_nlo\_as\_0119\_qed\_mc PDFs \cite{Ball:2012cx} & $\alpha_s(M_Z) = 0.119$ & $\alpha_{\rm EM} = 1/132.507$ \\ 
 & & \\
 \hline
\end{tabular}
\caption{Kinematics and input parameters used for the cross section computations in this paper. For our fixed-order computations in Section \ref{sec:fragmentation} we use the default scales $\mu_a=\mu_f=\mu_r=125\,{\rm GeV}$, where $\mu_r$ and $\mu_f$ are the renormalization and factorization scales, respectively, and $\mu_a$ is the scale associated with the non-perturbative fragmentation function. For the resummed results, we use $\mu_h=\mu_f=\mu_r= E^\gamma_T$, $\mu_j= R\, E^\gamma_T$ and $\mu_0= R \,E_0$ as the default.}\label{tab:input}
\end{table}

Let us now consider the inside fragmentation for fixed-cone isolation \eqref{eq:fixed}. This case is more complicated because the isolation fragmentation also involves non-perturbative fragmentation, see \eqref{eq:nonpertFrag}. At zeroth order in $\alpha_s$, there are two contributions. We can either have a trivial perturbative part
\begin{equation}
    \mathcal{I}_{i\to j}(z,R,E_\gamma, E_{\rm in}, \mu) = \delta_{ij}\, \delta(1-z) + O(\alpha_s)
\end{equation}
together with a non-perturbative fragmentation contribution, or we have photon production from a quark or anti-quark in the perturbative part $\mathcal{I}_{i\to \gamma}$ followed by the trivial photon-to-photon fragmentation $ \mathcal{D}_{\gamma \to \gamma}=\delta(1-z)$. Up to corrections of order $\alpha_s$, we can thus write the inside part for fixed-cone isolation as
 \begin{multline}\label{eq:nonpertFragNLO}
    \mathcal{F}^{\rm in}_{i\to \gamma}(z,R, E_\gamma,\, E_0,\mu) =\left[\mathcal{D}_{i \to \gamma}(z,\mu)
    + \sum_{k=q,\bar{q}}\,\delta_{ik} \,\mathcal{I}^{\rm in}_{k\to \gamma}(z, R, E_\gamma, \mu)\right] \theta\!\left(z-\frac{1}{1+\epsilon_\gamma}\right)
\end{multline}
and for the perturbative part, we find
\begin{equation}
   \mathcal{I}^{\rm in}_{q\to \gamma}(z, R, E_\gamma, \mu)= \frac{\alpha_{\text{EM}}\, Q_q^2 }{2 \pi} \left\{ P(z) \left[ -\frac{1}{\epsilon}   + \ln\!\left( \frac{\delta^2 Q^2}{\mu^2} (1-z)^2 z^2\right)\right] +z \right\}\,.
\end{equation}
Note that this is the opposite of $\mathcal{F}^{\rm out}_{q\to \gamma}$ in \eqref{eq:Fout}. In the absence of the isolation energy constraint in \eqref{eq:nonpertFragNLO}, the two contributions would exactly cancel since the perturbative part of the fragmentation function becomes scaleless. This is sensible: without the energy constraint, the isolation becomes trivial and the entire fragmentation reduces to the non-perturbative fragmentation function $\mathcal{D}_{i \to \gamma}$. We also note that the anomalous dimension of the fragmentation function is the same for smooth-cone and fixed-cone isolation. Since the same anomalous dimension also drives the evolution of the hard part given by the partonic amplitudes $d\sigma_{i + X}$ in \eqref{factorizationFormulaSmallCone} it cannot depend on the isolation requirement.

\begin{figure}[t!]
\centering
\begin{tabular}{cc}
\includegraphics[width=0.45\linewidth]{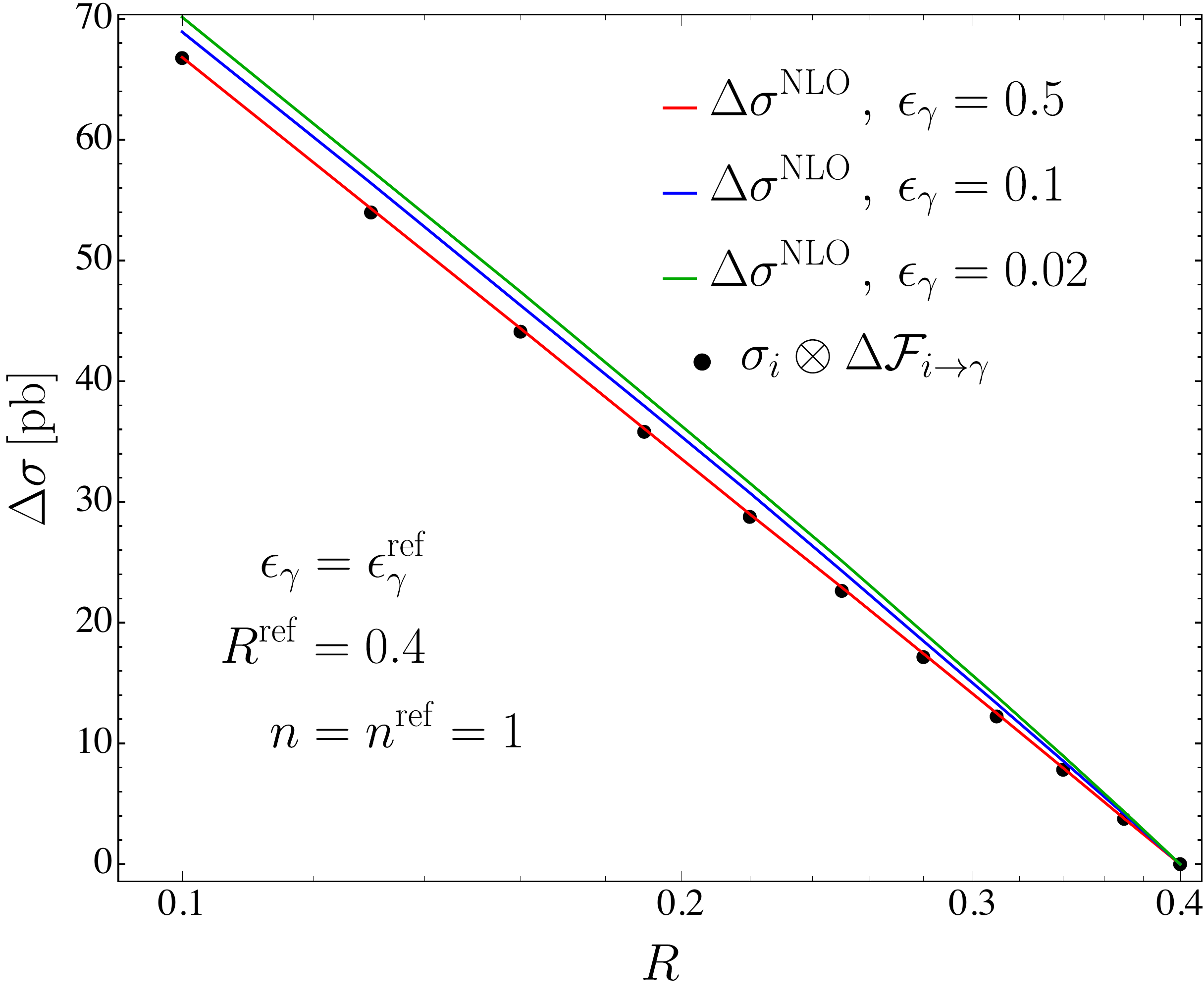} &
\includegraphics[width=0.45\linewidth]{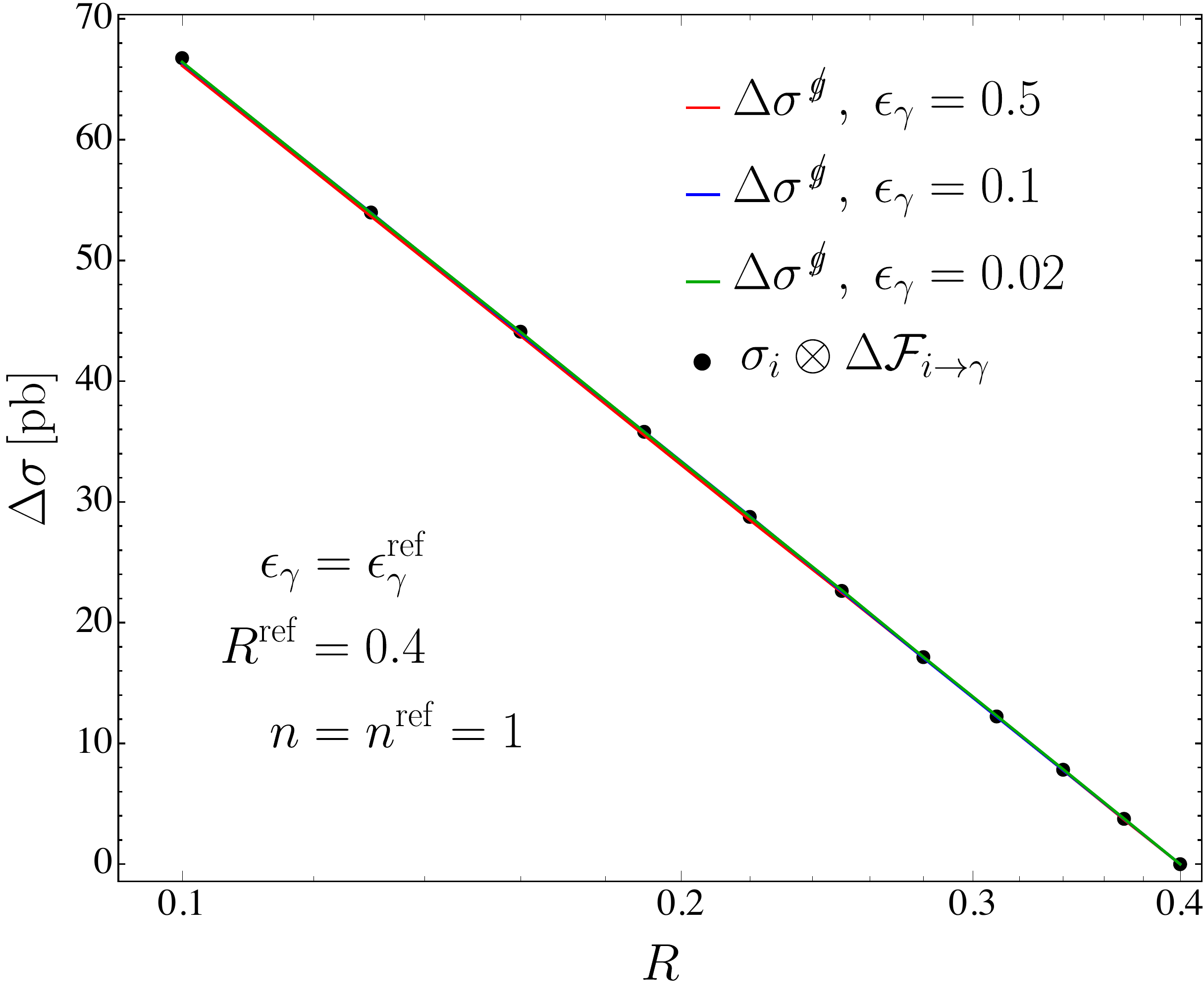}
\end{tabular}
\vspace{-2ex}
\caption{Dependence of $\Delta\sigma$ on the cone radius $R$ for smooth-cone isolation \eqref{eq:smooth}. The lines labeled $\Delta \sigma^{\rm NLO}$ are the difference of the full NLO cross sections.  For the lines labeled $\Delta \sigma^{g\hspace{-0.13cm}/}$ in the right plot gluons inside the isolation cones were vetoed. The dots represent $\Delta \sigma=\sigma_i \otimes \Delta\mathcal{F}_{i \to \gamma}$ computed with the fragmentation function according to \eqref{eq:diffFrag}, which is independent of $\epsilon_\gamma$ for $\epsilon_\gamma=\epsilon_\gamma^{\rm ref}$.}
\label{fig:R_dep}
\end{figure}

The fragmentation function factorization is valid up to power corrections in $R$ and with the functions at hand, it is interesting to check numerically whether \eqref{factorizationFormulaSmallCone} describes the isolation effects in the NLO photon production cross section at the experimentally used value $R=0.4$. To this end, we consider proton proton collisions at $\sqrt{s} = 13\,{\rm TeV}$ and compute the cross section for isolated photons with $E_T^\gamma > E_T^{\rm min}= 125\,{\rm GeV}$. For our numerical studies of photon isolation effects, we will use the kinematic setup and input parameters listed in Table \ref{tab:input} throughout the paper. Our formalism can also be used to study differential distributions, but the focus of our paper is on the effects of photon isolation and these are not strongly dependent on the photon kinematics.

To study the dependence on isolation parameters, we consider smooth-cone isolation \eqref{eq:smooth} and compute the difference to a reference cross section
\begin{equation}
    \Delta \sigma  = \sigma \left( \epsilon_{\gamma}, n , R  \right) - \sigma  ( \epsilon_{\gamma}^{\re},n^{\re}, R^{\re} )\,.
\end{equation}
In the difference $ \Delta \sigma$ the direct photon part in \eqref{factorizationFormulaSmallCone} drops out so that it is given by a convolution of the partonic cross section with the fragmentation function. At this order, the fragmenting parton is either a quark or anti-quark so that we have
\begin{equation}\label{eq:diffFrag}
\Delta \sigma = \sum_{i=q,\bar{q}}\, \int_{E_T^{\rm min}}^\infty dE_i    \int_{z_{\rm min}}^1 dz \frac{d\sigma_{i+X}}{dE_i} \Delta \mathcal{F}_{i \rightarrow \gamma}\, ,
\end{equation}
where 
\begin{equation}
\Delta \mathcal{F}_{i \rightarrow \gamma} =  \mathcal{F}_{i \rightarrow \gamma}  \left( z,R,\epsilon_{\gamma},n\right)  -  \mathcal{F}_{i \rightarrow \gamma} \left(z,R^{\re} ,\epsilon_{\gamma}^{\re},n^{\re} \right) \,
\end{equation}
and $z_{\rm min} = E_T^{\rm min}/E_i$. 

\begin{figure}[t!]
\centering
\begin{tabular}{cc}
\includegraphics[width=0.45\linewidth]{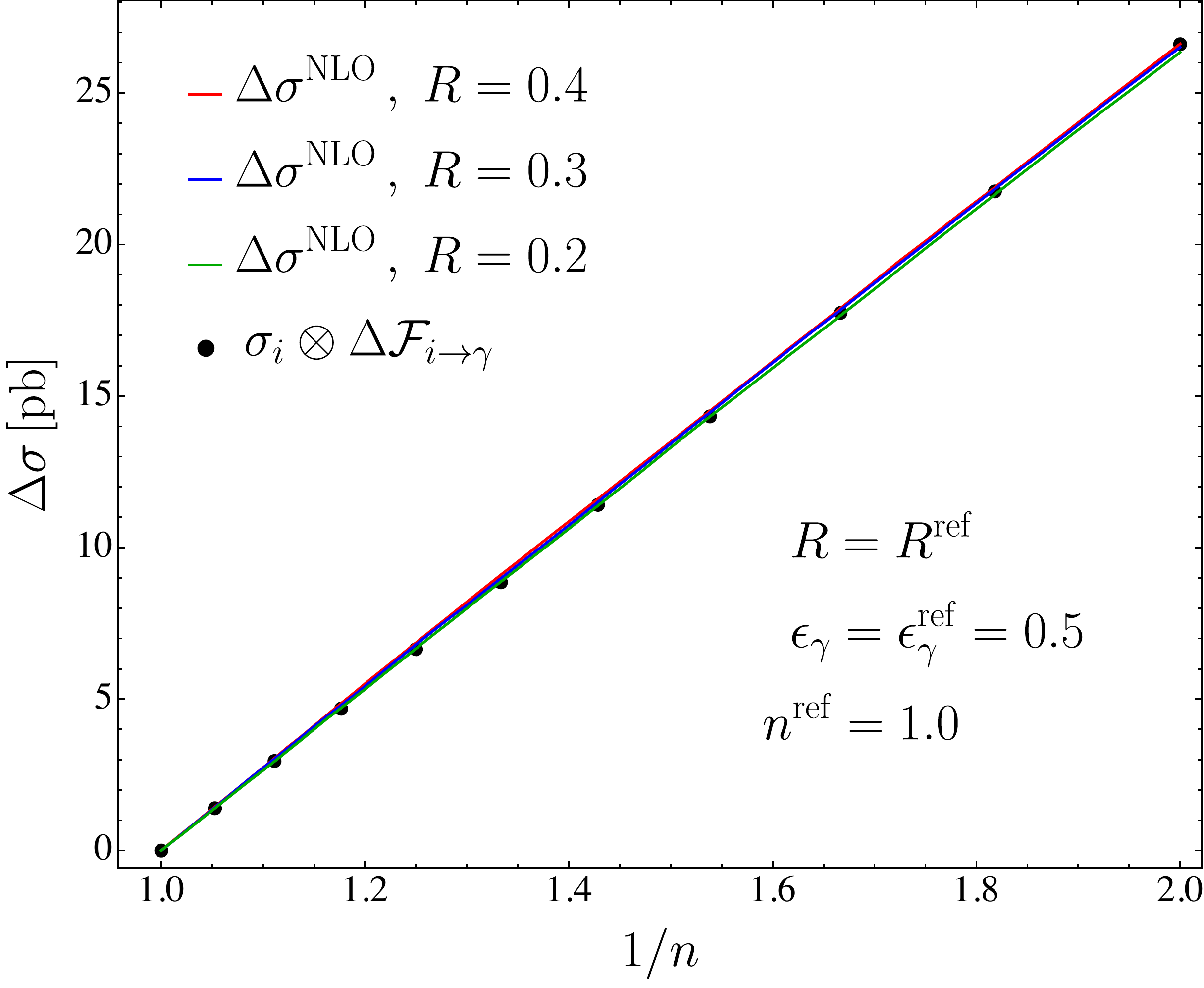} &
\includegraphics[width=0.45\linewidth]{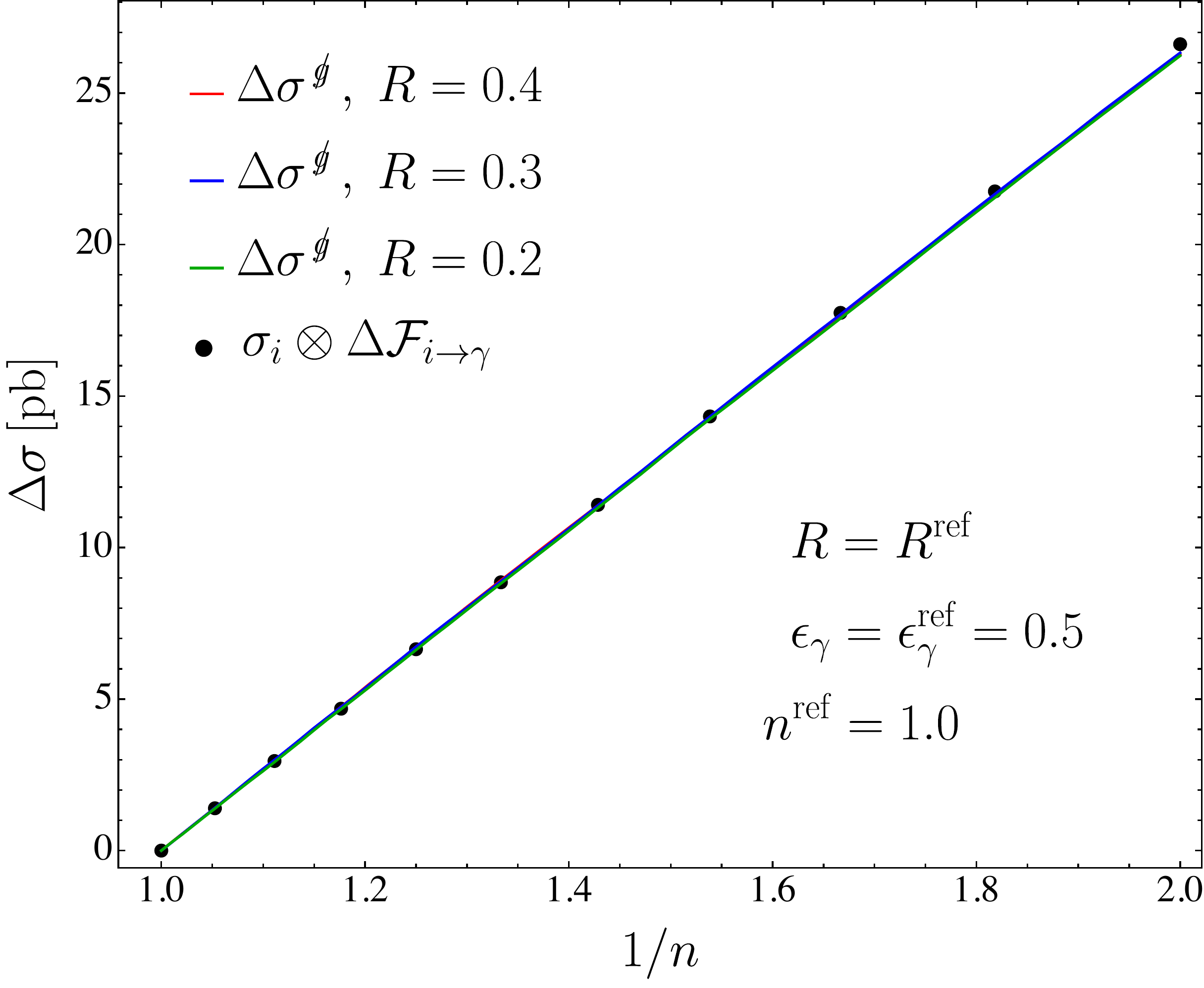}
\end{tabular}
\vspace{-2ex}
\caption{Dependence of $\Delta \sigma$ on the value of the parameter $n$ of the smooth-cone isolation \eqref{eq:smooth}. The lines labeled $\Delta \sigma^{\rm NLO}$ are the difference of the full NLO cross sections.  For the lines labeled $\Delta \sigma^{g\hspace{-0.13cm}/}$ in the right plot gluons inside the isolation cones were vetoed. The dots represent $\Delta \sigma=\sigma_i \otimes \Delta\mathcal{F}_{i \to \gamma}$ computed with the fragmentation function according to \eqref{eq:diffFrag} and is independent of $R$ for $R=R^{\rm ref}$.}
\label{fig:n_dep}
\end{figure}

To be able to convert the values for $\Delta \sigma$ into results for the full cross section, we computed some reference cross section values with {\sc MCFM} \cite{Campbell:2019dru} for the kinematics listed in Table \ref{tab:input}. The LO cross section is of course independent of the isolation requirement and corresponds to
\begin{equation}
\sigma^{{\rm LO}} = 229^{  -20}_{+22}\,{\rm pb} \,,
\end{equation}
where the upper and lower values correspond to the change in cross section after increasing and lowering $\mu_f = \mu_r$ from the default value by a factor 2, respectively. The NLO cross section values depend on isolation and we obtain
\begin{align}\label{eq:sigrefF}
\sigma^{{\rm NLO}} \Big |_{\text{no isolation}} &= 495^{-51}_{+68} \,{\rm pb}\,, \nonumber\\
\sigma^{{\rm NLO}} \Big |_{\text{fixed cone}, R=0.4 , \epsilon_\gamma = 0.02} &= 413^{-35}_{+46}  \,{\rm pb}\,, 
\end{align}
The cross section predictions in \eqref{eq:sigrefF} depend on the non-perturbative fragmentation functions $\mathcal{D}_{i \to \gamma}$ and we used the GdRG \cite{Gehrmann-DeRidder:1997fom,Gehrmann-DeRidder:1998bju} set as implemented in {\sc MCFM}. The code offers a second choice, the BFGS sets \cite{Bourhis:1997yu}, which would lead to a value of the cross section without isolation which is about $35 \,{\rm pb}$ lower. With fixed-cone isolation, the BFGS cross section would be $11 \,{\rm pb}$ lower than the value in \eqref{eq:sigrefF}. These fragmentation function sets were determined about 25 years ago based on {\sc LEP} data \cite{ALEPH:1995zdi,OPAL:1997lep} and models of the non-perturbative physics. For smooth-cone isolation, we obtain the reference values
\begin{align}\label{eq:sigrefS}
\sigma^{{\rm NLO}} \Big |_{R=0.4, \,n=1,\,\epsilon_\gamma = 1.0\phantom{0}} &= 459^{-43}_{+56} \,{\rm pb} \,,  \nonumber\\
\sigma^{{\rm NLO}} \Big |_{R=0.4, \,n=1,\,\epsilon_\gamma = 0.5\phantom{0}} &= 445^{-40}_{+53} \,{\rm pb} \,,\\
\sigma^{{\rm NLO}} \Big |_{R=0.4, \,n=1,\,\epsilon_\gamma = 0.02} &= 414^{-35}_{+46} \,{\rm pb} \,.  \nonumber
\end{align}
As a consistency check, we have computed cross sections with several available NLO codes and for convenience we provide precise reference values in Appendix \ref{app:ref}. We have also extracted the direct cross section in \eqref{factorizationFormulaSmallCone} by computing the cross section at different $R$ values, subtracting the fragmentation contribution and extrapolating to $R\to 0$. For the default scales in Table \ref{tab:input}, we find $\sigma^{\rm NLO}_{\rm dir}\approx 308 \,{\rm pb}$, with some uncertainty due to the extrapolation since we cannot run the fixed order codes at too small $R$ due to numerical instabilities.

From our results for the fragmentation function in \eqref{eq:FinSmooth} and \eqref{eq:Fout}, we can immediately read off the parameter dependence for a number of special cases, for example
\begin{align}
\Delta \sigma & \propto \ln\bigg(\frac{R^{\re}}{R}\bigg)  & &\text{ for} & &\; \text{$n=n^\re$  and $\epsilon_\gamma=\epsilon_\gamma^\re$}\,,  \nonumber\\
  \Delta \sigma  &\propto \left(\frac{1}{n} - \frac{1}{n^\re}\right)    & &\text{ for } && \text{ $R=R^\re$  and $\epsilon_\gamma=\epsilon_\gamma^\re$}\,.
\end{align}
In addition to the $n$ and $R$ dependence, we can also analyze the $\epsilon_\gamma$ dependence, but this case is more complicated because the difference of fragmentation functions has nontrivial dependence on the parameter $\epsilon_\gamma$ even for  $R=R^\re$ and $n=n^\re$:
\begin{equation}\label{eq:deltaFepsgamma}
\Delta \mathcal{F}_{i \rightarrow \gamma}=   \frac{ \alpha_{\rm EM} Q_i^2 }{2 \pi }P(z) \frac{1}{n} \left[\theta\!\left(z-\frac{1}{1+\epsilon_{\gamma} } \right)  \ln \left( \frac{1-z}{z\, \epsilon_{\gamma}}\right) -  \theta\!\left(z-\frac{1}{1+\epsilon_{\gamma}^{\re} } \right)  \ln \left( \frac{1-z}{z \,\epsilon_{\gamma} ^{\re} }\right)   \right]\,.
\end{equation}

\begin{figure}[t!]
\centering
\includegraphics[width=0.45\linewidth]{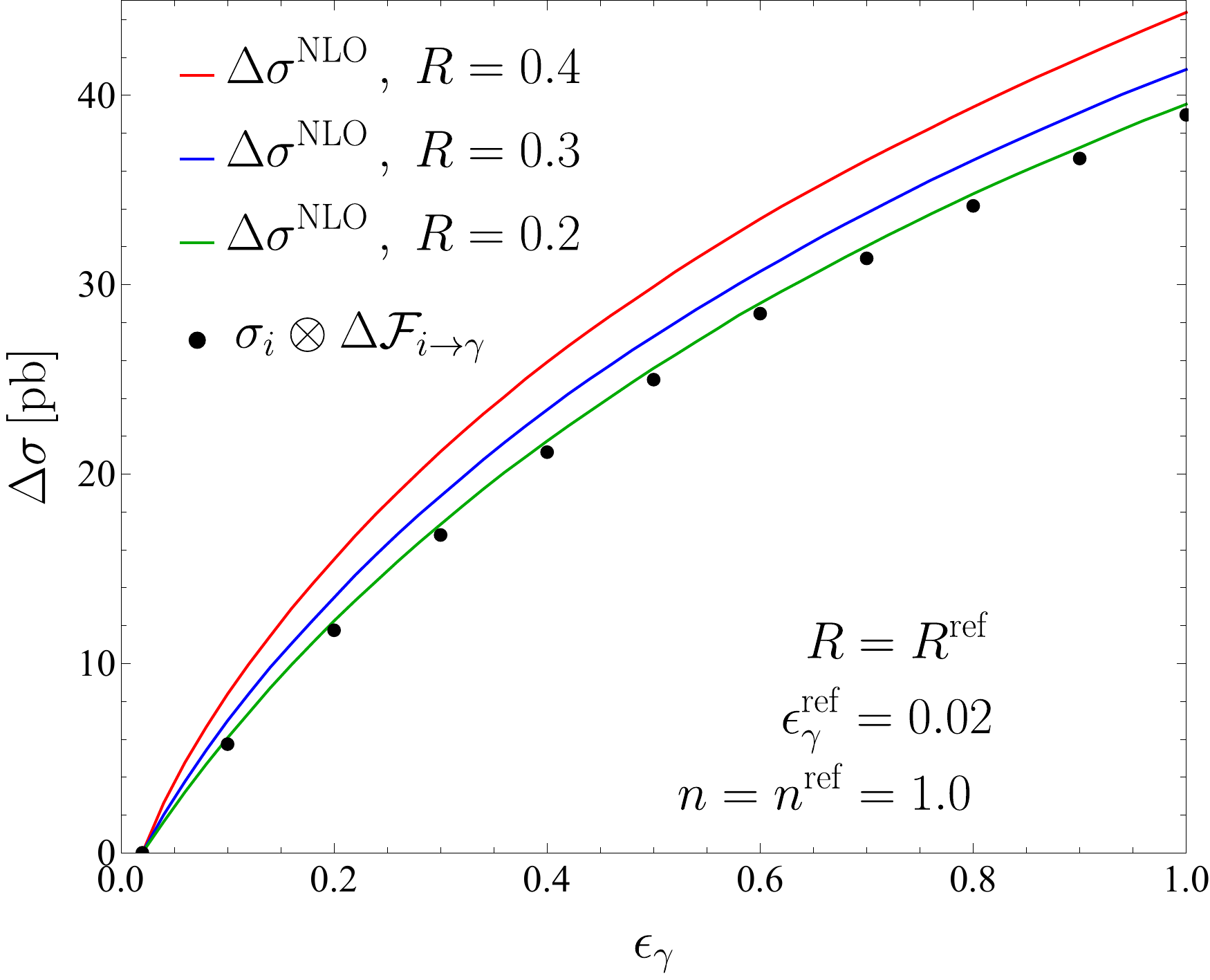}
\vspace{2ex}
\includegraphics[width=0.45\linewidth]{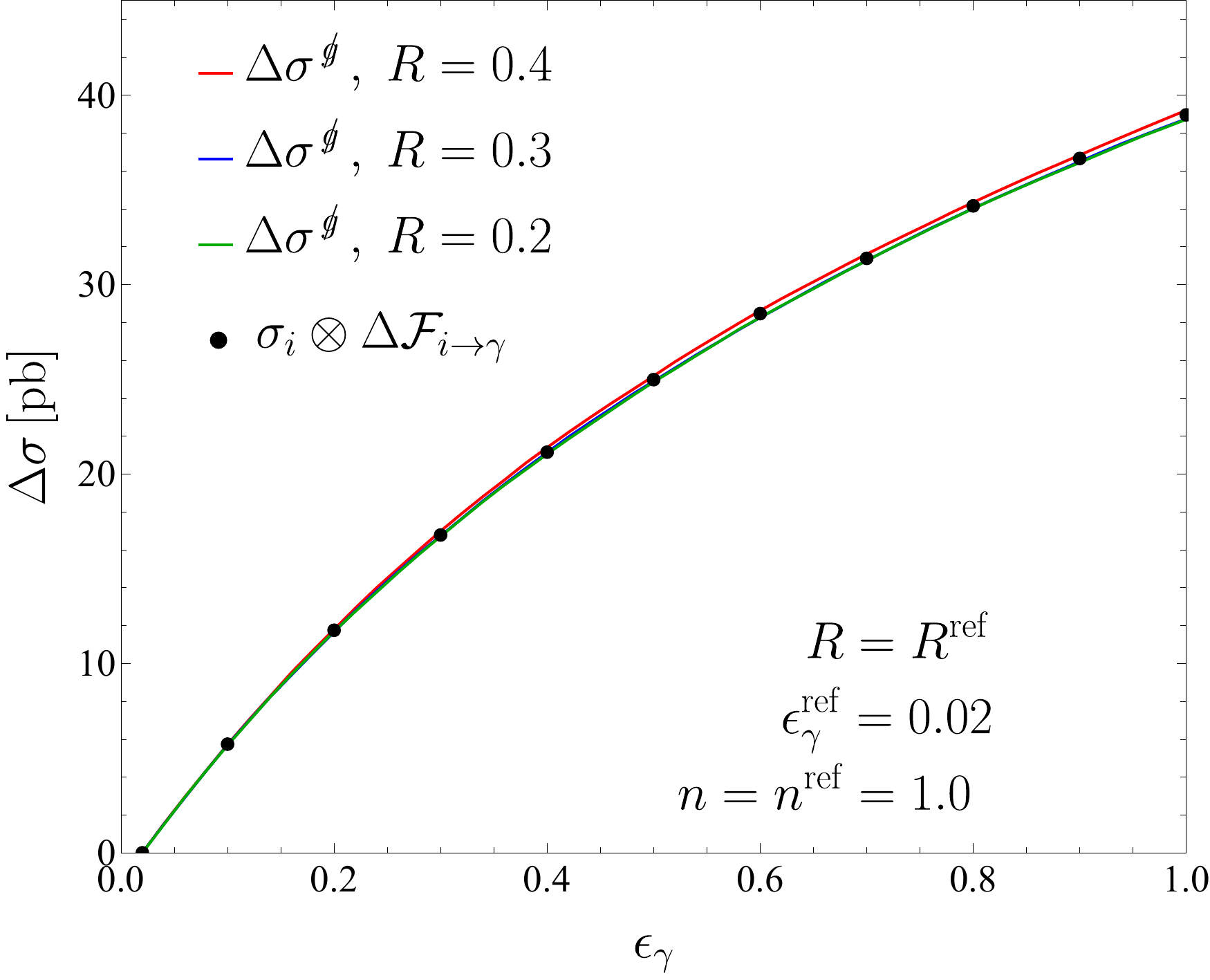}
\vspace{-2ex}
\caption{Dependence of the cross section on  $\epsilon_\gamma$ for smooth-cone isolation \eqref{eq:smooth}. The lines labeled $\Delta \sigma^{\rm NLO}$ show the difference of the full NLO cross sections. The dots represent $\Delta \sigma=\sigma_i \otimes \Delta\mathcal{F}_{i \to \gamma}$ computed with the fragmentation function according to \eqref{eq:diffFrag} and are independent of $R$ for $R=R^{\rm ref}$. For the lines labeled $\Delta \sigma^{g\hspace{-0.13cm}/}$ in the right plot gluons inside the isolation cones were vetoed.}
\label{fig:eps_dep}
\end{figure}

Of course, as is the case for the factorization formula \eqref{eq:diffFrag}, these results only hold up to terms which are power suppressed by the cone radius $R$ and it is interesting to check how big the power corrections are numerically by comparing to fixed-order results for $\Delta \sigma$.  To this end, we plot the cross section as a function the isolation parameters $R$, $n$ and $\epsilon_\gamma$ in Figures \ref{fig:R_dep}, \ref{fig:n_dep} and \ref{fig:eps_dep}. The dots in these figures correspond to fragmentation function results obtained using \eqref{eq:diffFrag}, while the lines are the NLO fixed-order result for $\Delta \sigma$ computed using {\sc MadGraph5\_aMC@NLO}  \cite{Alwall:2014hca}. The fixed-order photon production cross section only becomes sensitive to isolation at NLO and the cross section difference is insensitive to virtual corrections. We can thus extract the difference directly from a LO computation of the process $p p \to \gamma j j $, where one of the ``jets'' is recoiling against the photon, while the second one is inside the isolation cone. The details of this fixed-order computation are described in Appendix \ref{app:dSigma}.

The differences between the full fixed-order results (lines in the plots) and the fragmentation result (dots) are due to power suppressed contributions such as initial state radiation into the cone. 
Figures \ref{fig:n_dep} and \ref{fig:eps_dep} show that even for $R=0.4$, the power corrections are numerically quite small and the factorization theorem \eqref{factorizationFormulaSmallCone} accurately describes the photon isolation effects. In Figure \ref{fig:R_dep} the difference is zero by construction at the reference point $R=R^{\rm ref}=0.4$. Since the power corrections vanish for $R\to 0$, the difference in this region arises from power corrections to the reference cross section with $R^{\rm ref}=0.4$. Since the fragmentation contribution can only have (anti-)quarks inside the cone at this order, contributions with gluons inside the cone are suppressed by $R$. In the right-hand plots in Figures \ref{fig:R_dep}, \ref{fig:n_dep} and \ref{fig:eps_dep}, we have removed the contributions of gluons inside the cone. The close agreement of the fragmentation result with the full fixed-order result shows that gluon radiation into the cone is the main source of power corrections. Indeed, since the power corrections are so small, once gluons are excluded from the isolation cone, the three lines in each plot overlap almost completely. 

 \begin{figure}[t!]
\centering
\includegraphics[width=0.45\linewidth]{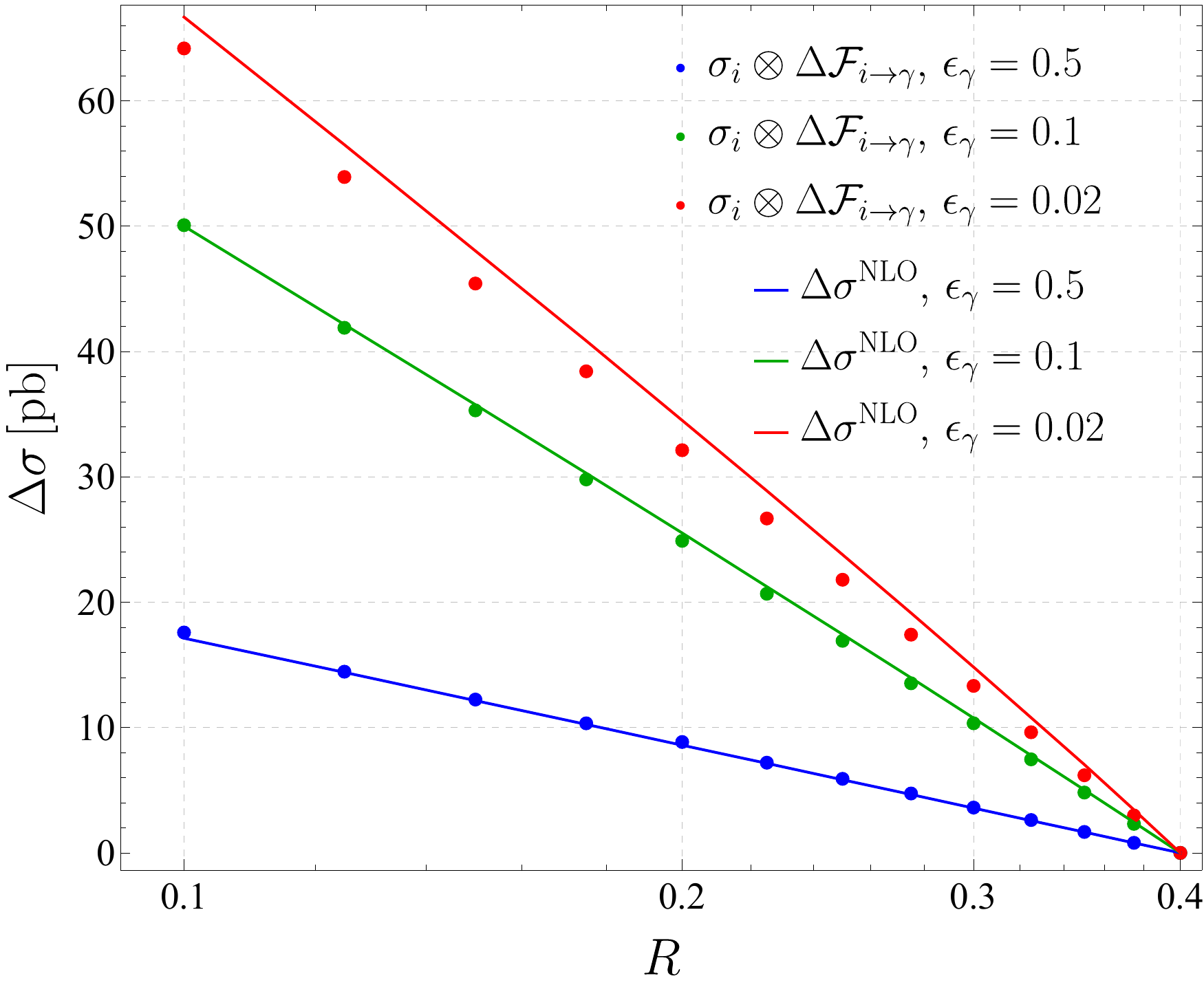}
\caption{Radius dependence for fixed-cone isolation for different $\epsilon_\gamma$ with $R^{\rm ref} =0.4$. The lines show the full NLO cross sections, the dots correspond the the result obtained using the cone fragmentation functions. In contrast to the smooth-cone result shown in Figure~\ref{fig:R_dep}, the result depends on $\epsilon_\gamma$.}
\label{fig:fixedRdep}
\end{figure}

In Figures \ref{fig:R_dep}, \ref{fig:n_dep} and \ref{fig:eps_dep} we  considered the parameter dependence of cross sections with smooth-cone isolation. It is now interesting to compare to the case of fixed-cone isolation. Since the outside part is obviously the same, different behavior is related to the inside part $\mathcal{F}^{\rm in}_{q\to \gamma}$ given in \eqref{eq:FinSmooth} and \eqref{eq:nonpertFragNLO}, respectively. In addition to the contribution from the non-perturbative fragmentation, a key difference between the two functions is that for fixed-cone isolation, the inside part of the function depends on the cone radius. Setting $\epsilon_ \gamma =\epsilon^{\rm ref}_ \gamma$ and computing the difference between the cross section at a given $R$ to a reference value $R^{\rm ref}$, the non-perturbative part drops out and we obtain
\begin{equation}
 \Delta  \mathcal{F}_{i \to \gamma} = \frac{Q_i^2  \alpha_{\rm EM}}{ \pi} P(z) \ln\!\left( \frac{R^{\rm ref} }{R} \right) \theta \!  \left( \frac{1}{1+\epsilon_{\gamma} } -z \right)\,.
 \label{eq:frag_fixed}
 \end{equation}
 We see that due to the presence of the $\theta$-function the coefficient of the logarithm of $R$ now depends on $\epsilon_\gamma$, in contrast to smooth-cone result shown in Figure~\ref{fig:R_dep}. The smaller the value of $\epsilon_\gamma$, the bigger the range over which the $z$-integral has support, resulting in a larger coefficient of the $\ln(R)$ term. This is indeed what we observe in Figure \ref{fig:fixedRdep}. In the limit $\epsilon_\gamma \to 0$, the $\theta$-function becomes trivial and we recover the smooth-cone result for the $R$ dependence of the cross section. This observation is surprising at first sight, but the underlying physics is easy to understand. For small $\epsilon_\gamma$, the $R$ dependence is driven by energetic partons outside the cone that are close to its boundary. These are independent of the isolation criterion so that the $\ln(R)$ dependence becomes universal. More generally, since the inside part $\mathcal{F}^{\rm in}_{i\to \gamma}$ involves a soft quark, its contribution is power suppressed for $\epsilon_\gamma \to 0$ and $R \to 0$. In this limit, a dependence on the isolation criterion first arises in the NNLO cross section and will be computed below. To illustrate that the different isolation criteria lead to the similar NLO cross section for $\epsilon_\gamma \to 0$, we have tabulated cross sections values for different isolation criteria in Table \ref{tab:MG__Xsec}. We  observe that the cross section differences indeed decrease for small $\epsilon_\gamma$. Interestingly, the $n=1$ cross section is fairly close to the fixed-cone cross section over a fairly wide range of $\epsilon_\gamma$ values.

Having illustrated the parameter dependence of the isolation cross section in different examples and demonstrated that power suppressed effects in $R$ are small, we now turn to the all-order resummation of $\ln(R)$ terms.

\section{Resummation of $\ln(R)$ terms}\label{sec:lnRres}

Working with the form \eqref{factorizationFormulaSmallConeAlt} of the factorization theorem, the renormalized fragmentation functions fulfills the usual DGLAP evolution equation
\begin{align}\label{eq:DGLAPhom}
\frac{d}{d\ln\mu} \mathcal{F}_{i\to \gamma}(z,\mu) &=  \sum_{j=\gamma,q,\bar{q},g} {\mathcal P}_{i\to j} \otimes \mathcal{F}_{j\to \gamma} \nonumber \\
& \equiv \sum_{j=\gamma,q,\bar{q},g} \int_z^1 \frac{dz'}{z'} {\mathcal P}_{i\to j}\Big(\frac{z}{z'}\Big) \mathcal{F}_{j\to \gamma}(z',\mu) \, ,
\end{align}
where we suppress the dependence on the fragmentation function on the additional arguments $E_\gamma$, $E_0$, $R$ and further parameters such as $n$.  As is conventional, we use here the symbol $\otimes$ to denote the Mellin convolution
\begin{equation}\label{MellColl}
(f \otimes g)(z) = \int_0^1 dx \int_0^1 dy \,\delta(z - x y) f(x) f(y) = \int_z^1\frac{dz'}{z'} f\Big(\frac{z}{z'}\Big) g(z')\,.
\end{equation}

\begin{table}[t!]
\centering
\begin{tabular}{c|c c c }
$\sigma$ [pb]  & fixed cone & $n=1$ & $n=2$ \\ 
\hline
$\epsilon_{\gamma} =0.02$  & $414.56 \pm 0.34 $ &  $413.31\pm 0.36 $ & $410.41\pm 0.37$\\
$\epsilon_{\gamma} =0.1$  & $ 420.58 \pm 0.38 $ &  $422.05 \pm 0.40 $ & $416.57 \pm 0.39$ \\
$\epsilon_{\gamma} =0.2$  & $429.35 \pm 0.32 $ &  $429.10\pm 0.41 $ & $421.71 \pm 0.40$ \\
\end{tabular}
\caption{Cross-section at $R=0.4$ for different photon isolation criteria computed using MCFM \cite{Campbell:2019dru}. The cross section values correspond to the kinematics and input specified in Table \ref{tab:input} with $\mu_f=\mu_r=125\,{\rm GeV}$. For this scale choice, the direct part of the cross section is $\sigma^{\rm dir}\approx 308\,{\rm pb}$.}
\label{tab:MG__Xsec}
\end{table}

Separating out the trivial $\mathcal{F}_{\gamma \to\gamma}$ contribution as in \eqref{eq:gammafrag}, we can write the DGLAP evolution equation purely in terms of QCD partons
\begin{align}
\frac{d}{d\ln\mu} \mathcal{F}_{i\to \gamma}(z,\mu) &= {\mathcal P}_{i\to \gamma}(z)+ \sum_{j=q,\bar{q},g} {\mathcal P}_{i\to j} \otimes \mathcal{F}_{j\to \gamma}\, ,
\label{eq:DGLAP}
\end{align}
with $i=q,\bar{q},g$. In this form, the equation involves an inhomogeneous term. To resum the logarithms of $R$ we will solve \eqref{eq:DGLAP} numerically and evolve the functions $\mathcal{F}_{i\to \gamma}$ from their characteristic scale $\mu_c \sim  R E_\gamma$ to the hard scale $\mu_h \sim E_\gamma$.  The initial condition $\mathcal{F}_{q\to \gamma}(z, E_\gamma,E_0, R,\mu)$ for $\mu=\mu_c$ was computed in the previous section both for fixed-cone and smooth-cone isolation. 

An important simplification for the case of smooth-cone isolation is that the fragmentation function is purely perturbative. The same is true in the limit of small $E_0$ considered in the next section, since the non-perturbative part involves a soft quark inside the fragmentation cone, which is power suppressed in the limit $E_0\to 0$. In the absence of non-perturbative effects, and since we do not include the top quark and set the masses of the other quarks to zero, our fragmentation functions have a flavor symmetry: all down-type quarks and anti-quarks have the same fragmentation function, and similarly all up-type quarks and anti-quarks. 
Instead of evolving the individual flavors, we thus only need the combinations
\begin{align}\label{eq:flavorC}
\Sigma&=\sum_{i=1}^{n_f} (\mathcal{F}_{q_i\to\gamma}+\mathcal{F}_{\bar{q}_i\to\gamma} ) \, , \nonumber\\
\Delta&= \mathcal{F}_{d\to\gamma}-\mathcal{F}_{u\to\gamma}
 \, , \\
 G&= \mathcal{F}_{g\to\gamma} \, , \nonumber
\end{align}
where $d$ and $u$ denote the down- and up-type quarks respectively. The function $\Delta$ is decoupled from the gluon fragmentation function, and satisfies the simple evolution equation
\begin{align}\label{eq:Delta}
\frac{d}{d\ln\mu} \Delta=\left( {\mathcal P}_{d\to \gamma}(z)-{\mathcal P}_{u\to \gamma}(z) \right)+ {\mathcal P}_{q\to q} \otimes \Delta  \, .
\end{align}
The other two functions $\Sigma$ and $g$ fulfill the coupled equations
\begin{align}\label{eq:Sigma}
\frac{d}{d\ln\mu}  \begin{pmatrix} \Sigma \\ G \end{pmatrix}=   \begin{pmatrix} \sum_{i=1}^{n_f} (\mathcal{P}_{q_i\to\gamma}+\mathcal{P}_{\bar{q}_i\to\gamma} ) \\ \mathcal{P}_{g \to\gamma} \end{pmatrix}+
\begin{pmatrix} {\mathcal P}_{q\to q} & 2n_f {\mathcal P}_{q\to g} \\ {\mathcal P}_{g\to q} &{\mathcal P}_{g\to g} \end{pmatrix} \otimes \begin{pmatrix} \Sigma \\ G \end{pmatrix}
  \, .
\end{align}
The parton-to-parton splitting kernels relevant for the homogenous part take the form
\begin{align}\label{eq:splitex1}
{\mathcal P}_{i\to j}(z)= \frac{\alpha_s}{\pi} P^{(1)}_{i\to j} +\mathcal{O}(\alpha_s^2) \, ,
\end{align}
and the parton-to-photon splitting kernels which constitute the inhomogeneous part of the equation are expanded as 
\begin{align} \label{eq:splitex2}
{\mathcal P}_{i \to \gamma}=\frac{\alpha}{\pi}\left( P^{(0)}_{i \to \gamma}+\frac{\alpha_s}{\pi}  P^{(1)}_{i \to \gamma} +\mathcal{O}(\alpha_s^2)  \right)\, .
\end{align}
We solve this equation at leading order in RG-improved QCD perturbation theory and therefore need to include the order $\alpha_s$ 
corrections to the evolution kernels, including the ones to ${\mathcal P}_{i \to \gamma}$. These can be found in \cite{deFlorian:2015ujt} and are listed in Appendix \ref{app:splitting}. In traditional terminology, this amounts to next-to-leading logarithmic (NLL) accuracy.

There are two commonly used techniques to solve evolution equations such as \eqref{eq:DGLAP}. One possibility is to solve the equations directly in momentum space by interpolating the fragmentation functions over a grid of $z$ values. In this approach computes the $\mu$-dependence step by step and interpolates the result in $z$ at each step. Alternatively, one can solve the equations in Mellin moment space 
\begin{equation}
f(N) = \int_0^1 dz z^{N-1} f(z)\,,
\end{equation}
which converts Mellin convolutions \eqref{MellColl} into products
\begin{equation}
(f \otimes g)(N) = f(N) \cdot g(N)\,.
\end{equation}
In moment space \eqref{eq:DGLAP} turns into a set of coupled differential equations for the moments. One can view the flavor indices as matrix indices so that the solution boils down to the solution of a matrix equation. The the inhomogeneous equation \eqref{eq:DGLAP} takes the form
\begin{align}\label{eq:DGLAPmom}
\frac{d}{d\ln\mu} \mathcal{F}_{i\to \gamma}(N,\mu) &=  {\mathcal P}_{i\to \gamma}(N) +  \sum_{j=q,\bar{q},g} {\mathcal P}_{i\to j}(N)  \mathcal{F}_{j\to \gamma}(N,\mu) \, .
\end{align}
Introducing the flavor combinations in \eqref{eq:flavorC}, we get a differential equation for $\Delta$ and a matrix differential equation for  $\Sigma$ and $G$, see \eqref{eq:Sigma}. After diagonalizing this two-by-two matrix, one can solve the equations analytically and obtain the exact $\mu$-dependence of the moments. The evaluation then reduces to computing the inverse Mellin transformation numerically. The moment-space solution is detailed in Appendix \ref{app:momentSol}.  The discussion in this appendix shows that for full NLL accuracy, one will need to include the two-loop correction to the parton-to-parton splitting kernels in \eqref{eq:splitex1} since $\mathcal{F}_{i\to \gamma}$ formally counts as $\mathcal{O}(1/\alpha_s)$. If the jet scale is not much lower than the hard scale, these corrections will be small and we omit them for simplicity.

\begin{figure}[t!]
\centering
\begin{tabular}{cc}
\includegraphics[width=0.45\linewidth]{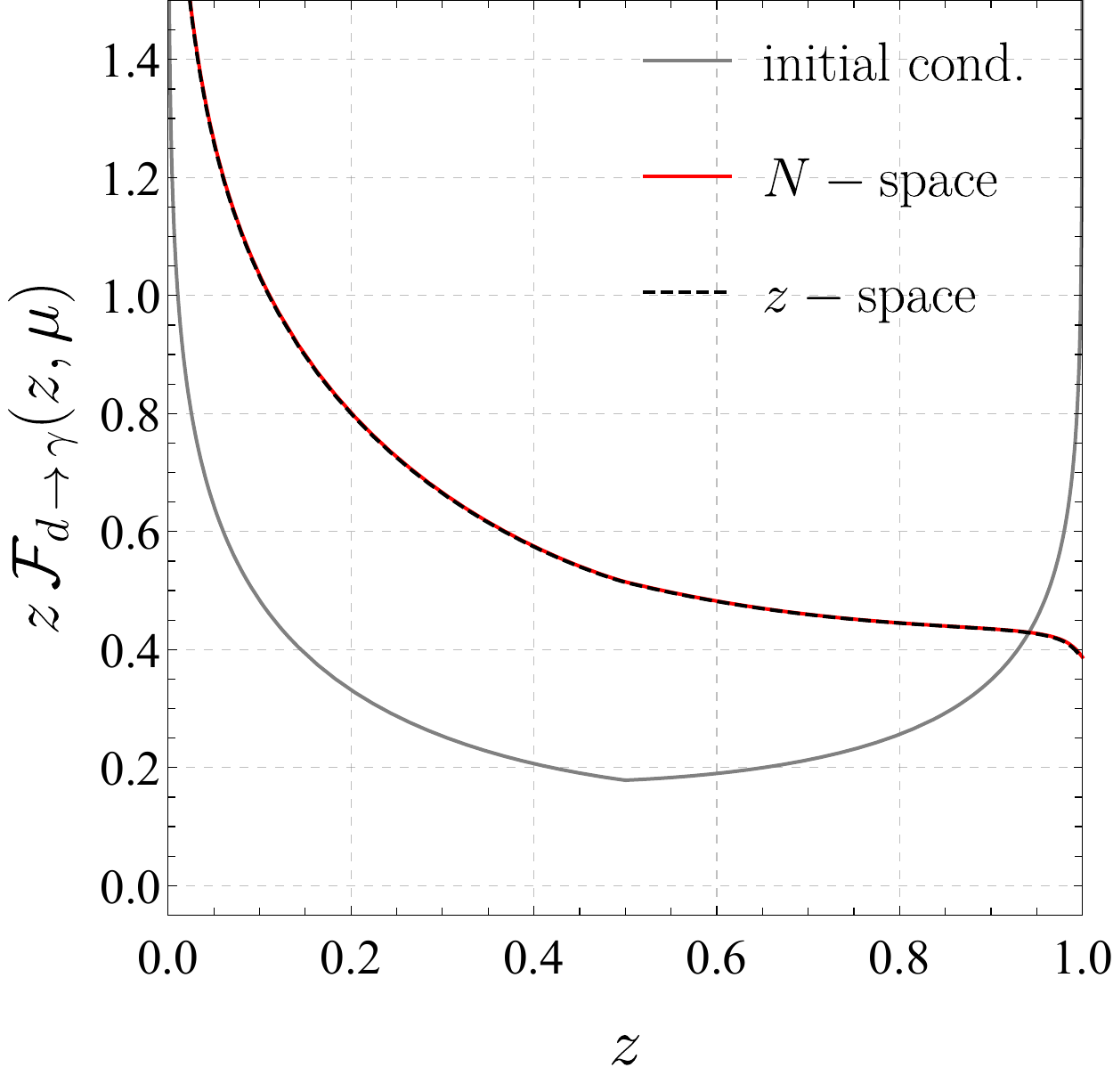} & \includegraphics[width=0.45\linewidth]{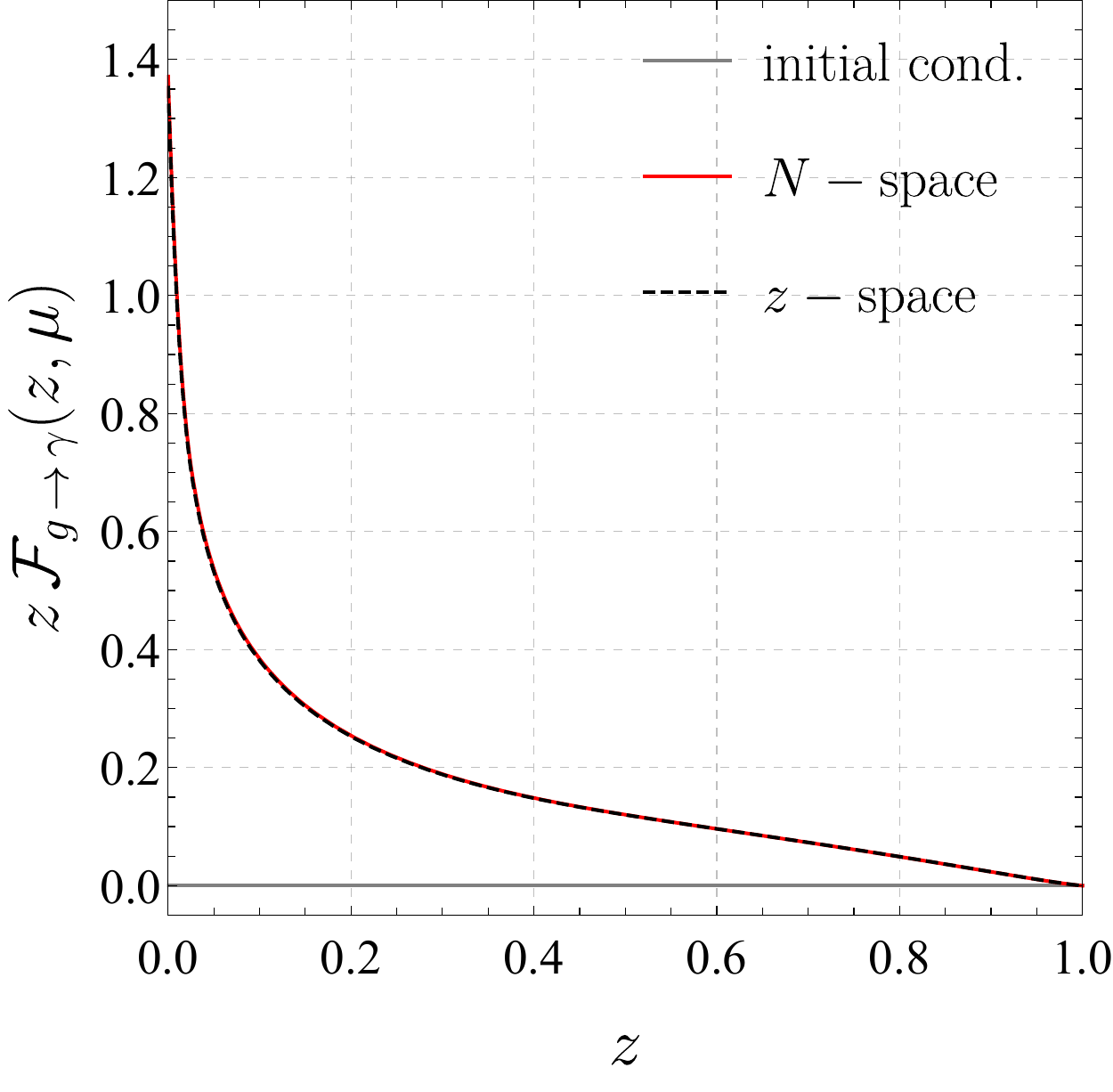}
\end{tabular}
\caption{Effect of RG-evolution on the cone fragmentation functions $\mathcal{F}_{i \to \gamma}$. The gray lines shows the initial condition given by the LO fixed-order result at $\mu = 10\, {\rm GeV}$ and correspond to smooth-cone isolation with $R=0.4$ with $\epsilon_\gamma = 1$ and $n=1$. The derivative of the initial condition is discontinuous at $z=(1+\epsilon_\gamma)^{-1} =0.5$ due to the contribution \eqref{eq:FinSmooth}. The gluon fragmentation function vanishes at this order. The other lines are the results after evolution to $\mu = 200\, {\rm GeV}$ by solving the RG equations either in moment space (red lines) or in momentum space (dashed lines).}
\label{fig:zFd}
\end{figure}

Both methods to solve the evolution equations are commonly used. The solution in moment space is, for example, the basis of the {\sc PEGASUS} code \cite{Vogt:2004ns}, while the {\sc APFEL} code solves the RGs in $z$-space \cite{Salam:2008qg,Bertone:2013vaa}. As a cross check, we have implemented both approaches. In Figure \ref{fig:zFd}, we compare results for some benchmark values of the scales and find that they are compatible with each other. The moment space method becomes delicate for $z\to1$ because the Mellin inversion integral suffers from slow numerical convergence. To improve the convergence, we use the same integration contour as the PEGASUS code. The momentum space method, on the other hand, requires a careful choice of the $z$ grid and interpolation and larger numerical resources to calculate the $\mu$-dependence since it needs to proceed in small steps, but yields similarly precise results for all $z$-values. In our event-based resummation framework, we prefer to work with the moment-space approach, since a single numerical integral immediately yields the result for any desired $\mu$ value. Of course, one could interpolate the results for the fragmentation functions as is done for PDFs, but one would need different grids for different initial conditions, i.e. different isolation parameter choices.

To compute the cross section resummed at NLL, we first evaluate the NLO photon-production cross section with {\sc MadGraph5\_aMC@NLO}  \cite{Alwall:2014hca}. Then we evaluate
\begin{equation}\label{eq:sigResumR}
\frac{d\sigma^{{\rm NLO+NLL}}}{dE_\gamma}
= \frac{d\sigma^{{\rm NLO}}_{\gamma+X}}{dE_\gamma}+\sum_{i=q,\bar{q},g} \int dz \frac{d\sigma_{i+X}}{dE_i} \Delta \mathcal{F}_{i\to \gamma} \,,
\end{equation}
where 
\begin{equation}
\Delta \mathcal{F}_{i\to \gamma}= \mathcal{F}_{i\to \gamma}(z, E_\gamma, E_0, R,\mu_j) -  \mathcal{F}_{i\to \gamma}(z, E_\gamma, E_0, R,\mu_h)\,.
\end{equation}
Here $\mu_h\sim E_\gamma$ is the scale at which the fixed-order computation was performed. The second term in $\Delta \mathcal{F}_{i\to \gamma}$ in \eqref{eq:sigResumR} subtracts the fixed-order result of the fragmentation contribution and adds the RG-improved result obtained from solving the evolution equation \eqref{eq:DGLAP} to evolve from the hard scale  $\mu_h$ at which $\sigma_{i+X}$ is computed  to the collinear scale $\mu_j \sim E_\gamma R$. The RG evolution resums the logarithms of $R$ and the subtraction is necessary to avoid a double counting of the fragmentation contribution which is contained in the NLO result. 

To compute the fragmentation contribution in \eqref{eq:sigResumR}, we use {\sc MadGraph5\_aMC@NLO}  as an event generator to produce the leading-order cross section $d\sigma_{i+X}/dE_i$ for different QCD partons $i$. We then perform the integral over $z$ in \eqref{factorizationFormulaSmallCone} by randomly choosing a value of $z$ for each event and evaluating the fragmentation function for this value. Since the scales depend on the photon energy $E_\gamma = z E_i$, we have a different $\mu$-values for each event and the moment-space technique to solve DGLAP is very efficient since we control the $\mu$ dependence analytically and only need a numerical integration to obtain the fragmentation function at the desired $z$ value. To have a fast way of computing the fragmentation function we have written a small C\texttt{++} code.
\begin{figure}[t!]
\centering
\includegraphics[width=0.7\linewidth]{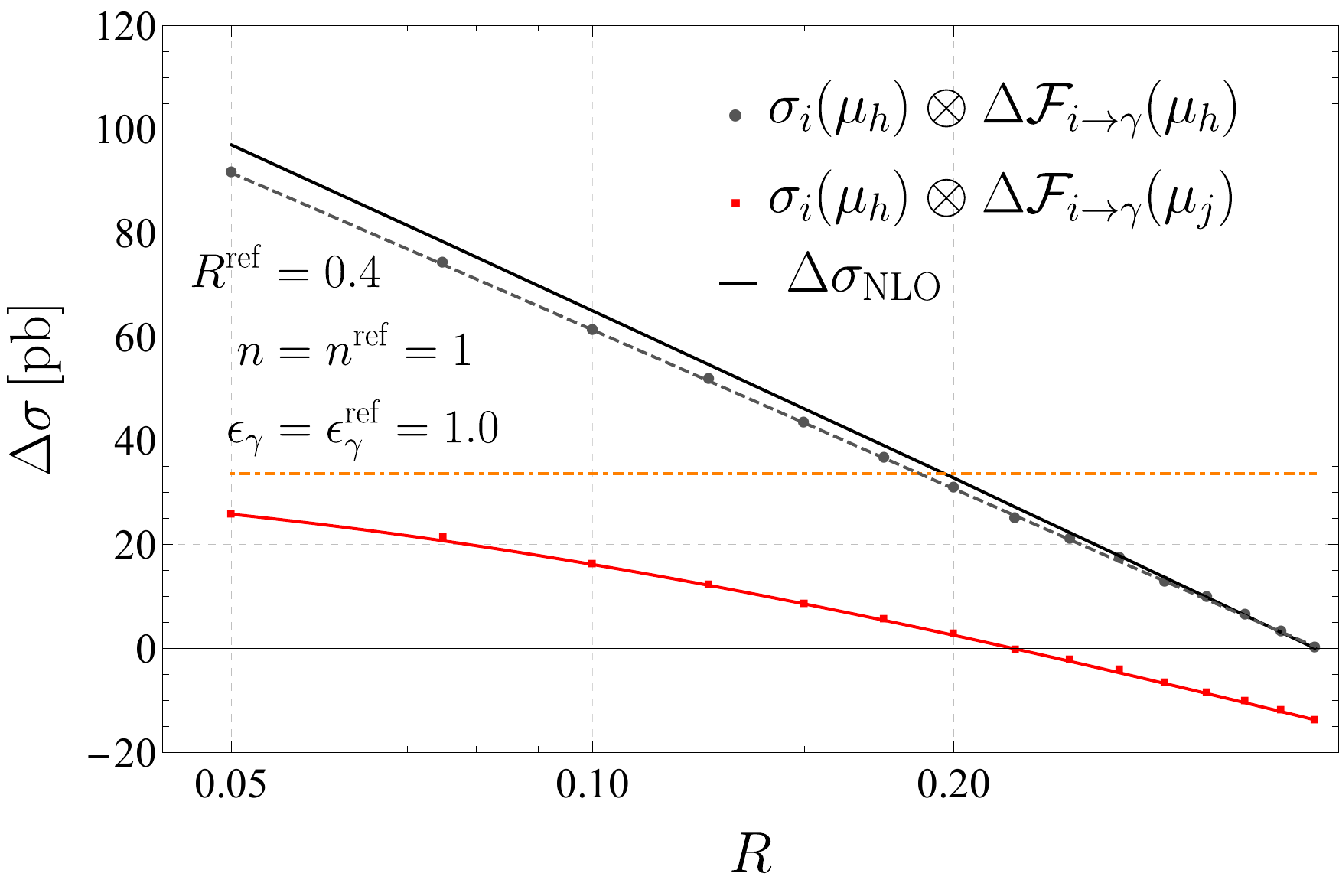}
\caption{Effect of $\ln(R)$ resummation, plotted as the difference to the fixed-order cross section at $R_{\rm ref} = 0.4$. Shown are the resummed result (red) and its fixed-order expansion (dashed) obtained by setting $\mu_j=\mu_h$. We also show the full fixed-order result (black) evaluated with $\mu_r=\mu_f=\mu_h$, which also includes terms which are power suppressed in $R$. Above the orange dot-dashed line, the cross section with isolation becomes larger than the inclusive cross section, which is unphysical.}
\label{fig:Delta_sigma_logR_resum}
\end{figure}

The effect of the resummation of the logarithms of the radius $R$ is shown in Figure~\ref{fig:Delta_sigma_logR_resum}. To show the dependence on $R$, we again compute the difference to a reference cross section at $R=0.4$. Before discussing the resummed result, let us compare the full NLO prediction (solid black line) of {\sc MadGraph5\_aMC@NLO} to the result obtained using the fragmentation formalism without resummation (dashed line). They must agree up to small power corrections and we observe that the difference is indeed quite small. Since we subtract the reference cross section, the difference is zero by construction at the reference point $R=0.4$. The small deviation at small $R$ is due to the difference of the reference cross sections as in Figures \ref{fig:R_dep} and \ref{fig:fixedRdep}. The red curve shows the difference of the resummed result to the reference cross section without resummation. As expected, resummation lowers the cross section since it dampens the logarithmic growth of the NLO result. We also show the difference between the inclusive photon cross section and the reference cross section obtained from \eqref{eq:sigrefS} as an orange, dash-dotted line in the figure. To be physical, the isolated cross section has to be smaller than the inclusive cross section. The fact that the isolated NLO cross section overshoots the inclusive result for $R < 0.2$ shows that the fixed-order expansion breaks down for small $R$, as was observed earlier in \cite{Catani:2002ny}. Resummation cures this problem. Of course, this unitarity bound has to be taken with a grain of salt, since the inclusive cross section depends on the non-perturbative fragmentation functions, which are poorly known.

\section{Factorization for small isolation energy $E_0$}\label{sec:smallIsolationEnergy}

If the isolation energy $E_0$ is much smaller than the photon energy $E_\gamma$, a scale hierarchy arises in the fragmentation function $\mathcal{F}_{i\to \gamma}$. In the limit of small $\epsilon_\gamma=E_0/E_\gamma$, energetic partons can no longer enter the isolation cone, however, energetic partons outside the cone can radiate back into the cone. This structure is at the heart of a second factorization, which is depicted in Figure \ref{fig:fact},
\begin{equation}\mathcal{F}_{i\to \gamma}(z, R \,E_\gamma, R\, E_0,\mu) = \sum_{l=1}^{\infty}\left\langle  \bm{\mathcal{J}}_{\!\! i \to \gamma+l}\!\left(\{\underline{n}\}, R \,E_\gamma,z,\mu \right)\otimes \bm{\mathcal{U}}_{l}\left(\{\underline{n}\}, \, R \,E_0,\mu\right)\right\rangle ,
\label{eq:deffrag}
\end{equation}
and is valid in the limit of small $\epsilon_\gamma$. The fragmentation function factorizes into jet functions $\bm{\mathcal{J}}_{\!\! i \to \gamma+l}$ describing the energetic partons accompanying the photon and functions $ \bm{\mathcal{U}}_{l}$ describing the low-energy radiation into the cone. This radiation is sensitive to the directions $\{\underline{n}\}= \{ n_1, \dots, n_l\}$ and color charges of the $l$ energetic partons. The symbol $\otimes$ denotes the integral over the directions of the hard partons and the photon. The same symbol was used in Section \ref{sec:lnRres} to denote the Mellin convolution; the context makes it clear what the symbol indicates. The notation $\langle \dots \rangle$ indicates the color sum, which can be taken after computing the emissions. In addition to directions of the $l$ energetic partons,  the functions also depend on the vectors $n$ and $\bar{n}$ introduced in defining  $\mathcal{F}_{i\to \gamma}$ and on the direction of the photon $n_\gamma$. More precisely, the functions will depend on scalar products of the different vectors, as we will detail below.

The fragmentation of parton $i$ into a photon of momentum $k$ is encapsulated by the jet functions
\begin{align}\label{jetfun}
   & \overline{\bm{\mathcal{J}}}_{\!i\rightarrow \gamma + l }(\{\underline{n}\},R \,E_\gamma,z,\mu) 
   = \sum_{\rm spins}  \prod_{j=1}^l \int \! \frac{dE_j \,E_j^{d-3} }{(2\pi)^{d-2}} \int \! \frac{dE_k \,E_k^{d-3} }{(2\pi)^{d-2}} \,    {\Theta }_{\rm cone}\!\left(\left\{\underline{p}\right\}\right)   \nonumber \\   & |\mathcal{M}_l(p_i;\{k,\underline{p}\})\rangle \langle\mathcal{M}_l(p_i;\{k, \underline{p}\})|
  2\,(2\pi)^{d-1}\, \delta(\bar{n}\cdot p_i - \bar{n}\cdot k-\bar{n}\cdot p_{X_c})\,\delta^{(d-2)}(k^\perp + p_{X_c}^\perp)\, ,
\end{align}
where the constraint ${\Theta }_{\rm cone}\!\left(\left\{\underline{p}\right\}\right)$ enforces that the energetic partons must lie outside the isolation cone. The amplitudes in this formula are the splitting functions for the incoming parton with momentum $p_i$ along the direction $n$ to fragment into the photon and additional $l$ energetic partons $\{\underline{p}\}= \{p_1, \dots, p_l\}$,
\begin{equation}
   |\mathcal{M}_l(p_i;\{\underline{p}\})\rangle
   = \langle k,\underline{p} |\,\Phi^{\alpha a}_c(0)\,|0\rangle \,,
\end{equation}
where $p_i=k +p_{X_c}=k +\sum_{j=1}^l p_j$ and $\Phi_c^{\alpha a}$ is a collinear field with the quantum numbers of the incoming parton, i.e. $\Phi_c^{\alpha a} = \chi_c^{\alpha a}$ for an incoming quark and $\Phi_c^{\alpha a} = \,\mathcal{A}^{\perp,\alpha a}_c$ for an incoming gluon, with spin and color indices $\alpha$ and $a$. The definition of the jet function includes a sum over spins of the outgoing partons, which for the quark-case produces the structure
\begin{equation}\label{eq:quarkJet}
 \overline{\bm{\mathcal{J}}}_{\!q\rightarrow \gamma +  l }(\{\underline{n}\},R \,E_\gamma,z,\mu) =  \left(\frac{n\!\!\!/}{2}\right)_{\alpha\beta} \delta^{ab} \bm{\mathcal{J}}_{\!q\rightarrow \gamma + l }(\{\underline{n}\},R \,E_\gamma,z,\mu)\,,
\end{equation}
with a scalar jet function $\bm{\mathcal{J}}_{\!i\rightarrow \gamma +  l }$,  and where $\alpha$ and $\beta$ are the Dirac indices of the collinear fields in the amplitude and the conjugate amplitude. For an incoming gluon, we instead get
 \begin{equation}\label{eq:glueJet}
\bar{n}\cdot p_i\,\, \overline{\bm{\mathcal{J}}}_{\!g\rightarrow \gamma + l }(\{\underline{n}\},R \,E_\gamma,z,\mu) =  - g^\perp_{\alpha\beta}\, g_s^2 \delta^{ab}  \, \bm{\mathcal{J}}_{\!g\rightarrow \gamma + l }(\{\underline{n}\},R \,E_\gamma,z,\mu)\,.
\end{equation}
The extra factor of $\bar{n}\cdot p_i$ on the left hand side arises because the gluon field has mass dimension $1$, while the quark field has dimension $\frac{3}{2}$. While we integrate over the full phase space of the photon with momentum $k$, the directions of the energetic partons are fixed. Note that the collinear fields in the jet functions scale as \eqref{pertColl}. The integrals in \eqref{jetfun} are integrals over the large light-cone components $E_j \equiv \bar{n}\cdot p_j/2$.
 
 The energetic partons in the jet functions source soft radiation which can enter the isolation cone. The momenta of this radiation scale as
\begin{equation}
( n \cdot p_t, \bar{n} \cdot p_t, p_t^\perp) \sim E_0 ( R^2, 1, R)\,.
\end{equation}
It has small energy $E \sim \epsilon_\gamma E_\gamma$ and is collinear to the photon. Since it is both collinear and soft it was called {\em coft} in \cite{Becher:2015hka} and denoted with a subscript $t$. The coft radiation can be obtained by taking matrix elements of Wilson line operators along the directions $n_1, \dots, n_l$ of the outgoing collinear partons and an additional one along the direction $\bar{n}$, which captures the radiation of all other partons not collinear to the photon. A detailed derivation of the Wilson line structure can be found in \cite{Becher:2016mmh}.
 The operator definition for the coft functions reads
\begin{multline}\label{eq:Um}
 \bm{\mathcal{U}}_l(\{\underline{n}\},E_0 R) \\= \int\limits_{X_t}\hspace{-0.58cm} \sum\,
   \langle 0|\, \bm{U}_0^\dagger(\bar{n})\,\bm{U}_1^\dagger(n_1)\dots {\bm U}_l^\dagger(n_l) 
   \, |X_t\rangle \langle X_t| \,\bm{U}_0(\bar{n})\dots {\bm U}_l(n_l)\, |0\rangle\, 
   \theta(2 E_0-\bar{n}\cdot p^{\rm cone}_{{X_t}})\,. 
\end{multline}
In the limit under consideration, the energy measurement translates into a measurement of the large component of the radiation. Note that the coft radiation can be inside or outside the cone, but only the energy of the partons inside the cone is bounded by $E_0$. 

In the following, we will resum the leading logarithms associated with the scale ratios shown on the right-hand side of Figure \ref{fig:scales}. The resummation of logarithms of the cone radius $R$, the ratio of the collinear scale $R E_\gamma$ to hard scale $E_\gamma$ is carried out as before by numerically solving the DGLAP evolution equation. However, for small $\epsilon_\gamma = E_0/E_\gamma$, a second evolution step is required to evolve from the collinear scale $\mu_j \sim R E_\gamma$ to the low-energy scale $\mu_0 \sim R E_0$ to resum the logarithms of $\epsilon_\gamma$. These are non-global logarithms and we resum them using a parton-shower algorithm \cite{Balsiger:2018ezi,Balsiger:2020ogy} analogous the one originally proposed by Dasgupta and Salam \cite{Dasgupta:2001sh}. 

\section{Computation of the jet and coft function\label{sec:jet}}

As a starting point of the second evolution, we need to compute the jet functions at a scale $\mu =\mu_j \sim R E_\gamma$. There are no large logarithms for this scale choice and all higher-multiplicity jet functions are suppressed by powers of $\alpha_s$. We thus only need the case $l=1$, corresponding to the fragmentation process $q \to  \gamma+q$.

The jet functions depends on the light-cone reference vector $n_\mu$ along the direction of the parton that fragments into the jet, as well as a conjugate reference vector $\bar{n}^\mu$ with $n\cdot \bar{n} =2$. In addition, the jet functions will depend on the light-cone reference vectors of the collinear partons produced in the fragmentation. For the lowest-order fragmentation process $q \to \gamma +  q $ we need a reference vector $n_q$  for the final-state quark and a vector $n_\gamma$ for the photon. The scalar jet functions $\bm{\mathcal{J}}_{\!q\rightarrow l +\gamma}$ defined in \eqref{eq:quarkJet} will depend on scalar products of these reference vectors and to compute them, we introduce angular variables that are suited to the limit under consideration. A set of variables which scales as $\mathcal{O}(1)$ is \cite{Becher:2016mmh}
\begin{align}\label{eq:ang1}
\Theta_i &= \frac{1}{\delta} \sqrt{\frac{n \cdot n_i}{\bar{n}\cdot n_i}}\,,\\
\Phi_{ij} &= \frac{2}{\delta^2} 
\frac{n_i \cdot n_j}{\bar{n}\cdot n_i \, \bar{n} \cdot n_j}  \,. \label{eq:ang2}
\end{align}
The first set of variables measures the angle with respect to the axis $n$, the second one the angle between $i$ and $j$. In four dimensions, we have
\begin{align}
 \Theta_i &= \frac{1}{\delta} \tan\left(\frac{\theta_i}{2}\right)\,,\\
 \Phi_{ij} &= \Theta_i^2 + \Theta_i^2 - 2 \Theta_i \Theta_j \cos(\Delta\phi_{ij}) \,.
\end{align}
For the leading order fragmentation process $q\to \gamma + q $ these variables are not independent. Momentum conservation enforces  $\Delta\phi_{q\gamma} = \pi$ and we therefore have $\Phi_{q\gamma} =  (\Theta_q+ \Theta_\gamma)^2$. Transverse momentum conservation also relates the ratio of the two angles to the momentum fraction $z$ of the photon
\begin{equation}
 \frac{\Theta_\gamma}{\Theta_q} =\frac{z}{1-z} + \mathcal{O}(\delta^2).
\end{equation}
This implies that there is only single independent angular variable and for convenience we choose it as
\begin{equation}
\widetilde{\Theta} = \frac{1}{\sqrt{\Phi_{q\gamma}}} = \delta \cot \frac{\theta_{q\gamma}}{2} + \mathcal{O}(\delta^2)\,.
\end{equation}
Up to power corrections, we have $\widetilde{\Theta} \in [0, 1]$. The limit $\widetilde{\Theta}=1$ corresponds to the quark touching the cone, while $\widetilde{\Theta}=0$ corresponds to the configuration where the photon and the quark are back to back. To rewrite the angular convolution integral in this variable, we insert
\begin{equation}\label{eq:thbar}
1= \int_0^1 \!d\widetilde{\Theta}\, \frac{2}{\widetilde{\Theta}^3}  \,\delta\!\left( \widetilde{\Theta}^{-2} - \frac{2}{\delta^2} 
\frac{n_q \cdot n_\gamma}{\bar{n}\cdot n_q \, \bar{n} \cdot n_\gamma}\right)\,,
\end{equation}
into the original angular convolution, perform the angular integrals and write the result in the form
\begin{multline}\label{eq:angtothbar}
\bm{\mathcal{J}}_{\!\! q \to \gamma+q}\!\left(\{n_q,n_\gamma\}, R \,E_\gamma,z,\mu \right)\otimes \bm{\mathcal{U}}_{q}\left(\{n_q,n_\gamma \}, \, R \,E_0,\mu\right) \\
 = \int_0^1 \!d\widetilde{\Theta}\,  \bm{\mathcal{J}}_{\!\! q \to \gamma+q}\!\left(\widetilde{\Theta}, R \,E_\gamma,z,\mu \right)\, \bm{\mathcal{U}}_{q}\left(\widetilde{\Theta}, \, R \,E_0,\mu\right).
\end{multline}

To compute the jet function for the process $q \rightarrow \gamma(k)+ q(p) $, we split the momenta into their light-cone components and write
\begin{equation}
p^\mu = n\cdot p \frac{\bar{n}^\mu}{2} + \bar{n}\cdot p \frac{n^\mu}{2} + p_\perp^\mu
\end{equation}
and analogously for the photon momentum $k$.
We note that
\begin{equation}
\frac{2 p\cdot k}{\bar{n}\cdot p\, \bar{n}\cdot k }=  
\frac{2 n_q \cdot n_\gamma}{\bar{n}\cdot n_q \, \bar{n} \cdot n_\gamma} = \frac{\delta^2}{\widetilde{\Theta}^2} \, .
\label{eq:appr}
\end{equation}
According to the definition \eqref{jetfun}, the jet function only involves the energy integrals instead of full phase-space integrations, but in \eqref{eq:angtothbar} we carry out the angular integrals after inserting the $\delta$-function \eqref{eq:thbar}. Doing so, we recover full phase-space integrals for $k$ and $p$ together with the $\delta$-function constraint \eqref{eq:thbar} which keeps the angle between the quark and the photon fixed. This gives
\begin{align}\label{eq:jetLO}
&\mathcal{J}_{q\rightarrow \gamma + q}\!\left(\widetilde{\Theta}, R \,E_\gamma,z,\mu \right) \delta^{ab} \left (\frac{ n \slsh }{2}\right)_{\alpha \beta} = \int [d p] [dk] \langle 0 | \chi^b_{\beta}(0)| \gamma \!+ \!q \rangle \langle \gamma \!+ \!q | \chi^a_{\alpha}(0)| 0 \rangle \, \nonumber
\\
&\quad\quad\; (2 \pi)^{d-1}  \delta^{(d-2)}(\vec{p}_{\perp}+\vec{k}_{\perp}) \delta(\bar{n}(p+k)-\tilde{Q})  \delta\Big(z- \frac{\bar{n} k}{\tilde{Q}}\Big)\, \frac{2}{\widetilde{\Theta}^3} \delta \left(\frac{2 p\cdot k}{\delta^2 \bar{n}\cdot p\, \bar{n}\cdot k}- \frac{1}{\widetilde{\Theta}^2}\right)  
\,
\end{align}
where $\tilde{Q} = 2E_\gamma /z$ is the large light-cone component of the quark before fragmentation. The matrix element is the same we encountered in the computation of the fragmentation function and was given in \eqref{eq:splitting}. The only difference to the earlier computation of the fragmentation function is the angular constraint. For the fragmentation function, the quark could be either inside or outside the cone according to \eqref{eq:constraintq} and we integrated over its direction. The particles inside the jet function, on the other hand, are energetic and cannot be inside the isolation cone. Furthermore we need the result differential in the direction $\widetilde{\Theta}$ of the quark, because the soft radiation depends on it. After inserting \eqref{eq:splitting} into \eqref{eq:jetLO}, we can immediately carry out the integrations which leads to the result 
\begin{align}\label{eq:jetfunres}
&\mathcal{J}_{q\rightarrow \gamma + q}\!\left(\widetilde{\Theta}, R \,E_\gamma,z,\mu \right) =  \;\frac{ \mu^2 e^{\gamma_E}}{\Gamma(1-\epsilon)}  \frac{  Q_i^2 \alpha_{\text{EM}}}{\pi}  \frac{2-2z +(1-\epsilon)z^2}{z} \frac{  \left(\frac{\delta ^2 Q^2 (z-1)^2 z^2}{\tilde{\Theta }^2}\right)^{-\epsilon }}{  \tilde{\Theta } } \, \nonumber
\\
&\quad\quad=\frac{  Q_i^2 \alpha_{\text{EM}}}{2 \pi} \left[ P(z) \left( \frac{\delta(\widetilde{\Theta})}{\epsilon}   -\delta(\widetilde{\Theta}) \ln  \left( \frac{\delta ^2 Q^2}{\mu^2} (z-1)^2 z^2\right)  +2 \left[ \frac{1}{\widetilde{\Theta}}\right]_{+}\right) -z \delta(\widetilde{\Theta}) \right]\,.
\end{align}
The splitting kernel $P(z)$ was given in \eqref{eq:splittingkernel}. The renormalized jet function is obtained by dropping the divergent term in the second line.

With the jet function at hand, we can now obtain $\mathcal{F}_{i\to \gamma}$ at leading order  from \eqref{eq:deffrag} by convoluting with the trivial lowest-order coft function $\mathcal{U}_q = \bm{1}$:
\begin{align}
\int_{0}^{1}\! d\widetilde{\Theta}\, \mathcal{J}_{q\rightarrow \gamma + q}\!\left(\widetilde{\Theta}, R \,E_\gamma,z,\mu \right) &=\frac{  Q_i^2 \alpha_{\text{EM}}}{2 \pi} \left[ P(z) \left( \frac{1}{\epsilon}   - \ln  \left( \frac{\delta ^2 Q^2}{\mu^2} (z-1)^2 z^2\right) \right) -z \right] \, .
\end{align}
This result indeed agrees with $\mathcal{F}^{\rm out}_{q\to \gamma}(z,R\, E_\gamma)$ given in \eqref{eq:Fout}. In the limit of small $\epsilon_\gamma$ the inside part is power suppressed, since soft quarks are power suppressed compared to soft gluons.

To resum the leading non-global logarithms, it is sufficient to use the trivial LO coft function since the function evaluated  at $\mu=\mu_0$ is free of large logarithms. It is nevertheless useful to calculate the NLO function  $\bm{\mathcal{U}}_q^{(1)}$, relevant for the process $q\to  \gamma +q $ so that we have an analytic result for the one-loop logarithm and and an idea of the size of the non-logarithmic $\mathcal{O}(\alpha_s)$ corrections. The perturbative expansion of the coft functions takes the form
\begin{equation}
\bm{\mathcal{U}}_{l}\!\left(\{ \underline{n} \} , \, R \,E_0,\mu\right)  = \bm{1} + \frac{\alpha_s}{4\pi} \bm{\mathcal{U}}_l^{(1)}\!\left(\{ \underline{n} \} , \, R \,E_0,\mu\right) +\mathcal{O} \left( \alpha_s^2\right)
\end{equation}
and the NLO correction to the coft function for $q\to \gamma + q$ is obtained by computing the emission of a coft gluon into the isolation cone
\begin{equation}\label{eq:nlocoft}
\frac{\alpha_s}{4\pi} \bm{\mathcal{U}}_q^{(1)}\left(\widetilde{\Theta}, \, R \,E_0,\mu\right)= 2 g_s^2 C_F \bm{1}  \int [dk]   \frac{\bar{n} \cdot n_q}{\bar{n} \cdot k \, n_q \cdot k} \theta\!\left(\delta^2-\frac{2 n_\gamma \cdot k}{\bar{n}\cdot n_\gamma \bar{n}\cdot k }\right)   \theta\!\left( Q_0 - \bar{n}\cdot k \right) \,.
\end{equation}
The first $\theta$-function forces the emission to lie inside the cone, the second one restricts the energy, or more precisely the large component of the coft momentum. The expression \eqref{eq:nlocoft} is relevant for fixed-cone isolation. For smooth-cone isolation in the limit of small $\delta$ one replaces
\begin{equation}
\theta\!\left( Q_0 - \bar{n}\cdot k \right)  \to  \theta \!\left(Q_0 \left( \frac{2 n_\gamma \cdot k}{\delta^2 \bar{n}\cdot n_\gamma \bar{n}\cdot k } \right)^n  - \bar{n}\cdot k \right) 
\end{equation}
and identifies $Q_0 = 2 \epsilon_\gamma E_\gamma$. Note that the one recovers the fixed-cone isolation for $n=0$. If the coft gluon is outside the cone its energy is unrestricted leading to a scaleless integral. The squared amplitude is from the emissions from the Wilson line along the direction $n_q$ of the outgoing quark and the Wilson line along the $\bar{n}$ direction which represents the emission from the remaining hard partons in the event. Performing the integrations for smooth-cone isolation, expressing the bare coupling $g_s$ through the $\overline{{\rm MS}}$ coupling, and defining $\bm{\mathcal{U}}_q^{(1)}= \mathcal{U}_q^{(1)} \bm{1}$ , we obtain
\begin{equation}\label{eq:Uq1}
\mathcal{U}_q^{(1)}  = \frac{2 C_F}{\epsilon} \ln\!\left(1-\widetilde{\Theta}^{2}\right)- 2 C_F  \left[\ln\!\left(1-\widetilde{\Theta}^{2}\right) 2 \ln\!\left( \frac{Q_0\delta}{\mu} \right)  +\ln^2\!\left(1-\widetilde{\Theta}^{2}\right)+(1+2n) {\rm Li}_2\! \left( \widetilde{\Theta}^{2}\right)\right]\,.
\end{equation} 
The renormalized one-loop function is obtained by dropping the divergent part of this result. For $\mu \sim Q \delta$, the result contains a large logarithm $\ln(Q_0/Q) = \ln(\epsilon_\gamma)$. The rest of the terms enter at NLL.

Given the simple form of the coft function \eqref{eq:Uq1}, we can analytically evaluate the convolution with the leading jet function in \eqref{eq:jetfunres}. Note that $\mathcal{U}_q^{(1)}$ vanishes for $\widetilde{\Theta} = 0$. For this reason, only the plus-distribution part in \eqref{eq:jetfunres} contributes. Evaluating the angular integral, we obtain
\begin{equation}\label{eq:JtimesU}
\left\langle \bm{\mathcal{J}}_{\!\! q \to \gamma+q}\otimes \bm{\mathcal{U}}^{(1)}_{q} \right\rangle = \frac{Q_q^2 \alpha_{\rm EM}}{\pi} C_F P(z) \left[-\frac{\pi ^2}{6 \epsilon } +\frac{\pi ^2}{3} \ln \frac{Q_0 \delta}{\mu} - (2 n+3)\, \zeta_3  \right]\,.
\end{equation}
This convolution of the jet and coft function corresponds exactly to the situation depicted in Figure \ref{fig:fact}.

The result \eqref{eq:JtimesU} has a very important application. Consider two cross sections computed at small $\epsilon_\gamma$ but with the same cone radius $R$. The difference  \eqref{eq:diffFrag} is proportional to the difference of fragmentation functions. Since the fragmentation contribution as a whole is suppressed by $\mathcal{O}(\alpha_s)$, we only need the fragmentation function difference to $\mathcal{O}(\alpha_s)$ to evaluate $\Delta \sigma$. In the limit of small $\epsilon_\gamma$ the fragmentation functions factorize into jet and coft functions and only the coft functions are sensitive to the isolation requirements. Since the jet functions are independent of the isolation criterion, the only contribution to the difference of cross sections
\begin{equation}
\Delta \sigma = \sigma_{\rm fixed cone}(R,\epsilon_\gamma) - \sigma_{\rm smooth cone}(R,\epsilon_\gamma^{\rm ref}, n) \,, 
\end{equation} 
arises from \eqref{eq:JtimesU} and takes the extremely simple form
\begin{equation}\label{eq:sigconv}
\Delta \sigma = \sum_{i=q,\bar{q}}\, \int_{E_T^{\rm min}}^\infty dE_i    \int_{z_{\rm min}}^1 dz \frac{d\sigma_{i+X}}{dE_i} \frac{Q_q^2 \alpha_{\rm EM}}{\pi} \frac{C_F \alpha_s}{4\pi} \,P(z) \left[ \frac{\pi ^2}{3} \ln \frac{\epsilon_\gamma}{\epsilon_\gamma^{\rm ref}} +2 n\,  \zeta_3 \right] .
\end{equation}
This formula holds at NNLO up to corrections suppressed by powers of $R$ or $\epsilon_\gamma$. 
As we have shown in the earlier sections, even at $R=0.4$, power suppressed effects are numerically small and experimental measurements use small values of $\epsilon_\gamma$. For reference, Figure \ref{fig:fixedRdep} shows the size of the power suppressed effects at NLO. For $\epsilon_\gamma =0.02$, the $R$ dependence at NLO is indeed very close to the one for smooth-cone isolation. We also provided NLO cross section values in Table \ref{tab:MG__Xsec} to indicate the size of the remaining differences. Of course, to make optimal use of the formula \eqref{eq:sigconv}, one would only use it to convert the NNLO corrections and separately compute the NLO fixed-cone results so that the power suppressed corrections to the formula are also suppressed by $\alpha_s^2$. Numerically,  the value of $\Delta \sigma$ obtained from \eqref{eq:sigconv} is quite small. Computing it for our standard setup detailed in Table \ref{tab:input} for $n=1$ and $\epsilon_\gamma=\epsilon_\gamma^{\rm ref}$, we obtain
\begin{equation}
\Delta \sigma = -1.3\,{\rm pb}\,.
\end{equation}

\section{Resummation of $\ln(\epsilon_\gamma)$ terms\label{sec:resummation}}

\begin{figure}[t!]
\centering
 \includegraphics{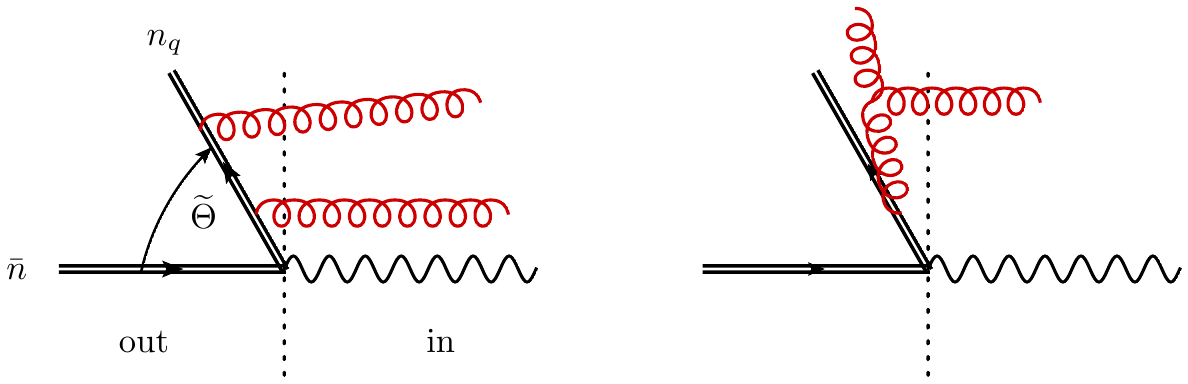} 
\vspace{-2ex}
\caption{Kinematics and example diagrams arising in the the parton shower computation of $\bm{\mathcal{U}}^{\rm LL}_{q}(\widetilde{\Theta}, t )$. We use the rescaling invariance \eqref{eq:rescale} and Lorentz invariance of the shower to evaluate the function in a frame where $\bar{n}$ and $n_\gamma$ are back-to-back and the isolation cone covers the entire right hemisphere. The left diagram shows an example of a two-loop global contribution, the right one a non-global one.}
\label{fig:shower}
\end{figure}

\begin{figure}[t!]
\centering
\begin{tabular}{ccc}
\includegraphics[height=0.45\linewidth]{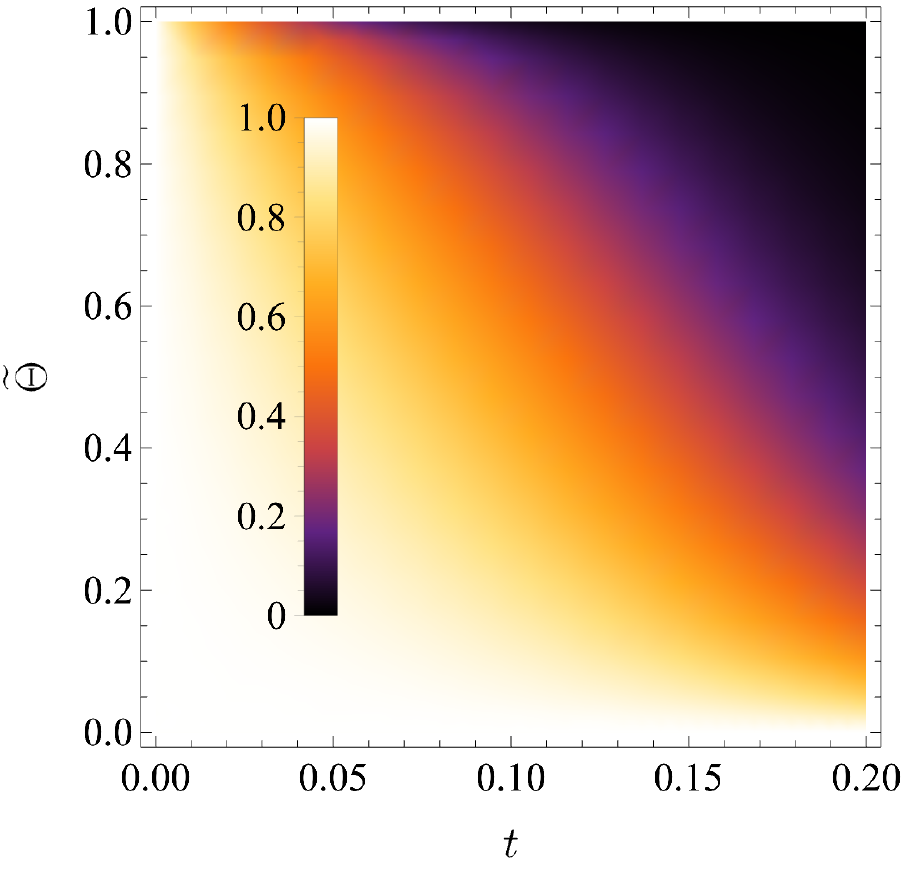} & & \includegraphics[height=0.45\linewidth]{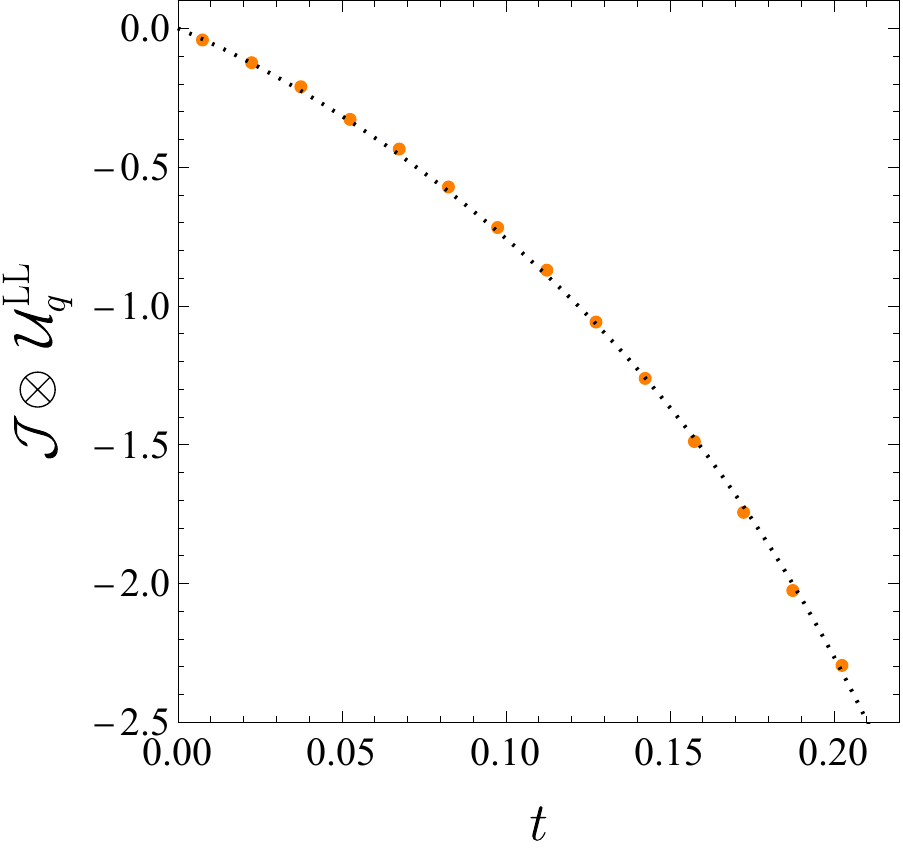} 
\end{tabular}
\vspace{-2ex}
\caption{The left plot shows the coft function $\mathcal{U}_q^{\rm LL}(\widetilde{\Theta},t)$. The right plot shows the convolution of the jet and coft function, more precisely the convolution with the plus distribution shown in \eqref{eq:PlusCoft}. The orange dots correspond to the results of the parton shower; the dotted line is the fourth-order polynomial in $t$ fitted to these results.} 
\label{fig:utheta}
\end{figure}

To resum the leading logarithms of $ \epsilon_\gamma$ we solve the RG equations and run the jet function from the jet scale $\mu_j\sim Q R$ down to the scale $\mu_0 \sim Q_0 R = Q \epsilon_\gamma R$, where we combine it with the coft functions. To perform the resummation we will use the parton-shower code {\sc NGL\_resum} \cite{Balsiger:2020ogy}. This code was developed to numerically perform the RG evolution and the angular integrals over the directions of the additional partons that are emitted during the evolution. It is not possible to apply the code directly to our problem, since we work in the limit $R \to 0$, where the size of the isolation region goes to zero and the MC integration over the angles would become highly inefficient, as additional emissions would be enhanced by logarithms of $\delta$. To use the code, we use the fact that the coft function is not depending on $\delta$ and $Q_0$ individually, but only on the product. This is explicit in the one-loop result \eqref{eq:Uq1}, but can be proven formally by noting that the coft function is invariant under the rescaling
\begin{equation}\label{eq:rescale}
\begin{aligned}
\delta &\rightarrow  \frac{ \delta}{\lambda}\,, &
Q_0 & \rightarrow \lambda\, Q_0\,, &
\bar{n} &\rightarrow \lambda\, \bar{n}\,,  &
n_{\gamma} &\rightarrow \frac{ n_{\gamma}}{\lambda} \,,&
n_{i} &\rightarrow \frac{ n_{i}}{\lambda} \, .
\end{aligned}
\end{equation}
To see this, note that the Wilson lines in \eqref{eq:Um} are invariant under rescalings of the light-cone vectors and the rescaling also leaves the constraints on the energy and the angular variables \eqref{eq:ang1} and \eqref{eq:ang2} invariant. Setting $\lambda = \delta$, the invariance implies that we can run the shower for $\delta = 1$, where the opening angle is $\pi/2$ after rescaling the energy to $Q_0 \delta$. The shower computes the leading-logarithmic (LL) evolution
\begin{equation}
\mathcal{U}^{\rm LL}_{q}\left(\widetilde{\Theta}, \, R \,E_0,\mu_0\right) =\sum_{m=2}^\infty  \Big\langle  \bm{U}_{2m}(\{\bar{n},n_q,\underline{n}\},\mu_j,\mu_0)\,\hat{\otimes}\, 
\bm{1} \Big\rangle \,,
\end{equation}
where the evolution matrix
\begin{align}
\bm{U}(\{\bar{n},n_q, \underline{n}\},\mu_j,\mu_0) = {\rm \bf P} \exp\left[ \int_{\mu_0}^{\mu_j} \frac{d\mu}{\mu} \bm{\Gamma}^{H}( \{\underline{n}\},\mu) \right], \label{eq:evolmat}
\end{align}
produces additional partons along the directions $\{\underline{n}\}$ and the symbol $\hat{\otimes}$ indicates the integral over their directions. For LL resummation the exponent of the evolution matrix reduces to
\begin{equation}\label{eq:evolutiontime}
\int_{\mu_0}^{\mu_j} \frac{d\mu}{\mu}\, \bm{\Gamma}^H= \int_{\alpha_s(\mu_0)}^{\alpha_s(\mu_j)} \frac{d\alpha}{\beta(\alpha)}\, \frac{\alpha}{4\pi}\,\bm{\Gamma}^{(1)}=  \frac{1}{2\beta_0}\ln\frac{\alpha_s(\mu_0)}{\alpha_s(\mu_j)} \,\bm{\Gamma}^{(1)}  \equiv t \,\bm{\Gamma}^{(1)}  \,.
\end{equation}
The ``evolution time'' $t$ measures the separation of the scales $\mu_j$ and $\mu_0$. The relevant one-loop anomalous dimension $\bm{\Gamma}^{(1)}$ can be found in \cite{Becher:2016mmh} (by now also the two-loop result is known \cite{Becher:2021urs}) and the solution of the evolution equation is detailed in \cite{Balsiger:2018ezi,Balsiger:2020ogy}. The shower starts with a quark along the direction $\bar{n}$ which fragments into a photon and a quark along the $n_q$ direction with angular separation $\widetilde{\Theta}_q$. We use that the shower is Lorentz invariant to choose a frame where
\begin{align}
n_\gamma^\mu & = (1,0,0,1)\,,  &\bar{n}^\mu &= (1,0,0,-1)\,, & n_q^\mu &= (1,\sin \theta_{\gamma q},0,\cos \theta_{\gamma q})\,,
\end{align}
so that $\widetilde\Theta = \cot(\theta_{\gamma q}/2)$ for our choice $\delta=1$. The shower then generates successive emissions outside the isolation cone (or hemisphere for $\delta=1$) , until one emission is inside the cone after which it terminates. The resulting function $\mathcal{U}^{\rm LL}_{q}(\widetilde{\Theta}, t )$ is plotted on the left hand side of Figure \ref{fig:utheta}. One thing that is obvious in the one-loop result \eqref{eq:Uq1} and in the plot is that the function is trivial for $\widetilde{\Theta}=0$, where 
$\mathcal{U}^{\rm LL}_{q}(\widetilde{\Theta}, t )=1$ independently of $t$. For $\widetilde{\Theta}=0$, the outgoing quark lies along the direction $\bar{n}$ of the fragmenting quark. In this configuration, the radiation of the two exactly cancels. One can view the fragmenting quark in the initial state as an anti-quark in the final state to make the cancellation manifest. The most radiation arises for $\widetilde{\Theta}=1$, which corresponds to a configuration where the quark is at the edge of the isolation cone.

 One can interpolate the function $\mathcal{U}^{\rm LL}_{q}(\widetilde{\Theta}, t )$ and then evaluate the angular convolution with the jet function \eqref{eq:angtothbar}, but it is more efficient to also Monte-Carlo integrate over $\widetilde{\Theta}$ and directly compute the convolution \eqref{eq:angtothbar} inside the parton shower code. To do so, we we first use $\mathcal{U}^{\rm LL}_{q}(0, t )=1$ to compute the $\delta$-function terms and obtain
\begin{multline}
\int_0^1 \!d\widetilde{\Theta}\,  \left\langle \bm{\mathcal{J}}_{\!\! q \to \gamma+q}\!\left(\widetilde{\Theta}, R \,E_\gamma,z, \mu_j \right)\, \mathcal{U}^{\rm LL}_{q}(\widetilde{\Theta}, t ) \right\rangle \\
 = \frac{  Q_i^2 \alpha_{\text{EM}}}{2 \pi} \Big[ -P(z)  \ln  \left(\frac{\delta ^2 Q^2}{\mu_j^2} (z-1)^2 z^2\right)  -z +  2 P(z) \int_{0}^{1} d\widetilde{\Theta} \left[ \frac{1}{\widetilde{\Theta}}\right]_{+} \mathcal{U}^{\rm LL}_q(\widetilde{\Theta},t)  \Big] \, .
\end{multline}
The variable $t\equiv t(\mu_j,\mu_0)$ encodes the dependence on the low scale. 
Since $\mathcal{U}^{\rm LL}_{q}(0, t )=1$, we can drop the plus-prescription and evaluate 
\begin{align}\label{eq:PlusCoft}
\int_{0}^{1} \frac{d\widetilde{\Theta}}{\widetilde{\Theta}}\, \mathcal{U}^{{\rm LL}}_{q}(\widetilde{\Theta}, t ) \approx -\frac{\pi^2}{2} t-31.5 t^2 +105.\, t^3  -535.\, t^4  \, .
\end{align}
where the result on the right-hand side was obtained by fitting a fourth order polynomial to the numerical parton shower results, obtained by sampling the $\widetilde\Theta$ integral with $25\times 10^5$ values and running $500$ showers at each value. The results of the parton shower are compared to the fit results in the right panel of Figure \ref{fig:utheta} and agree quite well up to values of $t$ relevant for phenomenological applications. Up to running effects, the different powers of $t$ correspond to the successive terms in the expansion in $\alpha_s$ since
\begin{equation}
t = \frac{\alpha_s}{4\pi} \ln \frac{\mu_j}{\mu_0} + \mathcal{O}(\alpha_s^2).
\end{equation}
 The leading term in the expansion in $t$ can be obtained analytically by integrating the logarithmic term in our result \eqref{eq:Uq1}. In addition to performing the resummation, our parton shower also computes this term and the numerical value agrees with the analytic result to an accuracy better than a permille. The term linear in $t$ is captured by NNLO fixed-order computations of photon production. NNLO is necessary since $\sigma_{i+X}$,  the cross section to produce the fragmenting parton, is $\alpha_s$ suppressed. Our parton shower also computes the coefficient of the $t^2$ term, with an accuracy of order of a few per cent. The remaining two terms were determined by fitting to the shower results. Taking the exponential of the one-loop contribution yields the ``global'' logarithms. Adopting this terminology,  the two-loop term is split into 
 \begin{equation}
-  31.5 = -43.7\, (\text{``non-global''}) +12.2\, (\text{``global''})
\end{equation}
where we approximated $ \pi^4/8 \approx 12.2$. The non-global part is thus significantly larger than the global part and the same remains true at higher orders. Diagrams for the two contributions are shown in Figure \ref{fig:shower}. The global contribution arises from emissions from the quark before or after the fragmentation as indicated on the left diagram  in Figure \ref{fig:shower}. The non-global terms arise from sequential emissions off gluons emitted outside the isolation cone, see the right diagram. In the shower, we include a collinear cutoff $\eta_{\rm cut}=6$ (see \cite{Balsiger:2018ezi,Balsiger:2020ogy})  as well as a technical cutoff $\widetilde{\Theta}>\kappa_{\rm cut}\approx 10^{-3}$ in the angular integral. We have checked that for $t< 0.2$ our results are insensitive to these cutoffs. The fact that we are able to fit the shower results with a polynomial in $t$ implies that the fixed-order expansion of the logarithmic terms is well-behaved in the region we perform our computation.  

\begin{figure}[t!]
\centering
\begin{tabular}{c}
\includegraphics[width=0.7\linewidth]{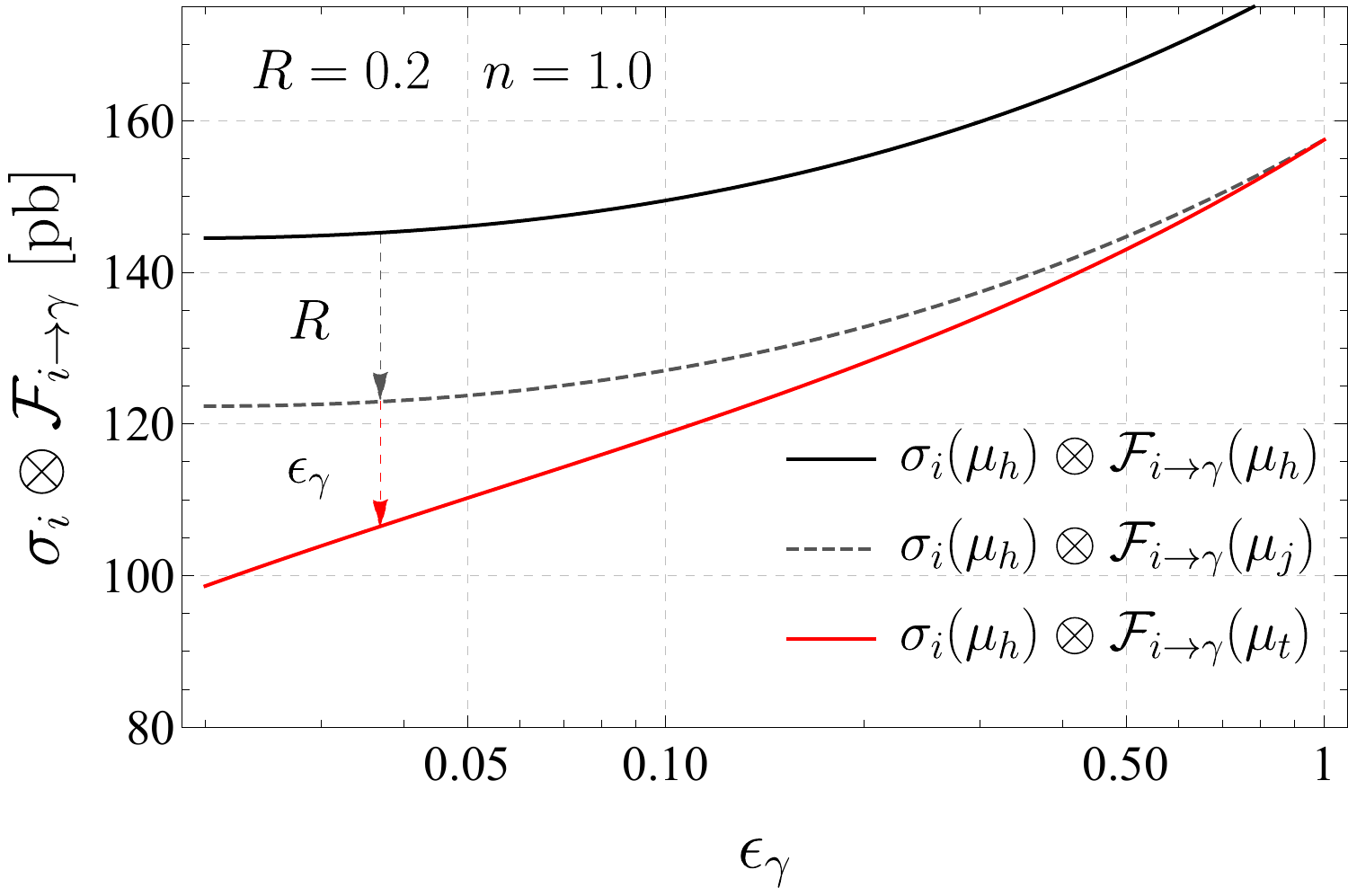} \\[10pt]
\includegraphics[width=0.7\linewidth]{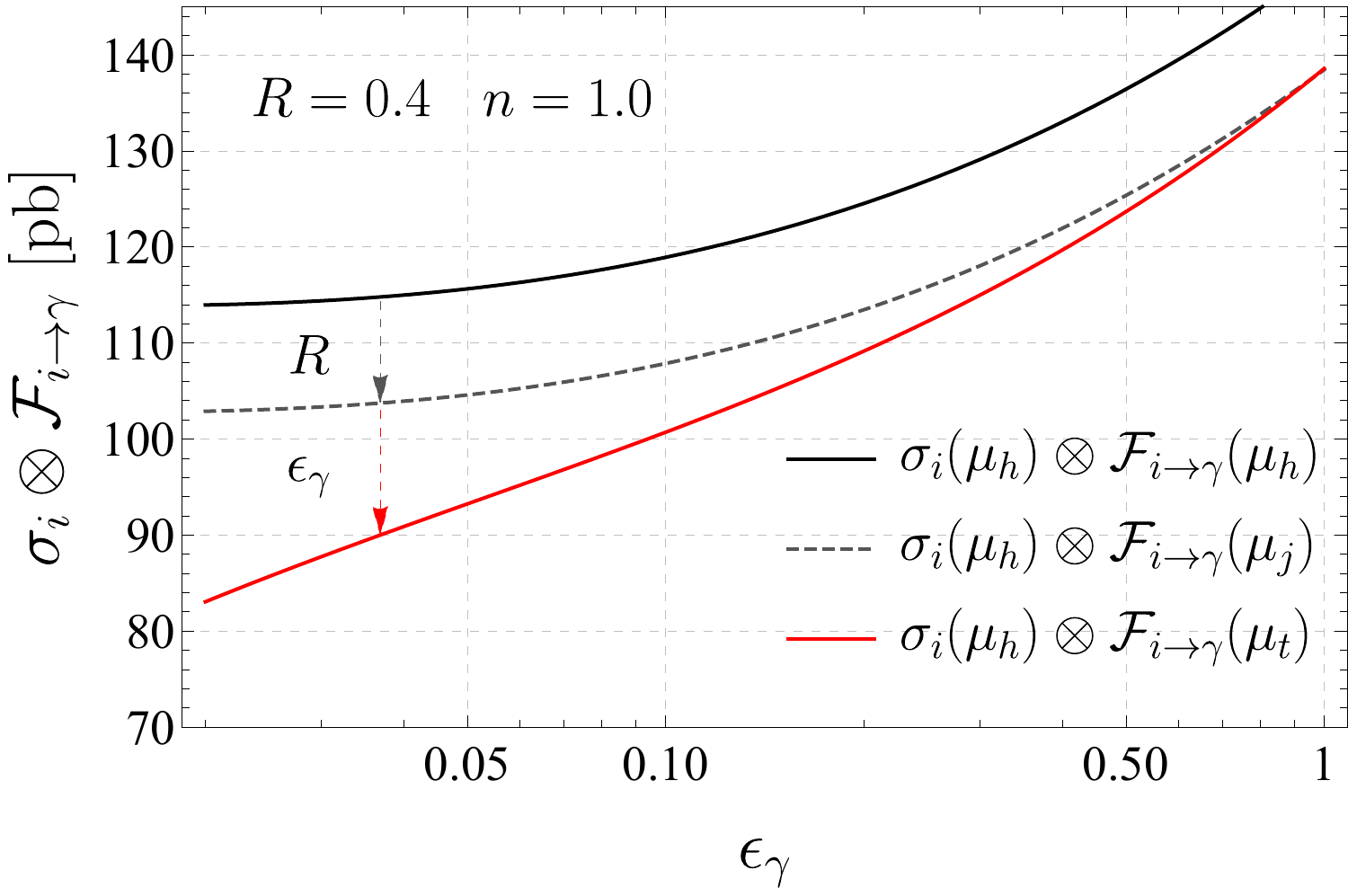} 

\end{tabular}
\vspace{-2ex}
\caption{Effect of the resummation of $\ln(R)$ and $\ln(\epsilon_\gamma)$ terms for $R=0.2$ (upper plot) and $R=0.4$ (lower plot). The solid black curve shows the result without resummation, the dashed curve includes the resummation of $\ln(R)$ terms. The red curve resums both types of logarithms. Only the fragmentation contribution is shown, to obtain the full cross section the direct photon production contribution with $\sigma^{\rm dir} \approx 290\,{\rm pb}$ has to be added. (The direct cross section is somewhat lower than the one given in Section \ref{sec:fragmentation} because we use dynamic scales rather than fixed ones, see Table \ref{tab:input}.)} 
\label{fig:lnepsResum}
\end{figure}

Let us now look at the effect of the resummation on the cross section. For illustration, we will again 
consider proton proton collisions at $\sqrt{s} = 13\,{\rm TeV}$ and compute the cross section for isolated photons with $E_T^\gamma > E_T^{\rm min}= 125\,{\rm GeV}$ and $|\eta_\gamma| < 2.37$ following {\sc ATLAS} \cite{Aad:2019eqv}. The result shown in Figure \ref{fig:lnepsResum} includes both the resummation of $\ln(R)$ and $\ln(\epsilon_\gamma )$ terms. The resummation is achieved by first evolving from the hard scale $\mu_h \sim E_\gamma$ to the jet scale $\mu_j \sim R \,E_\gamma $ by solving the DGLAP equation as discussed in Section \eqref{sec:lnRres} and then evolving to the coft scale $\mu_0 \sim   \epsilon_\gamma\, R \, E_\gamma $ using the parton shower framework. We can distinguish the effect of the two resummations by choosing different scales. Setting $\mu_j = \mu_h$ switches off the $\ln(R)$ resummation, while the choice $\mu_j = \mu_0$ eliminates the higher-order $\ln(\epsilon_\gamma )$ terms. The effect of these choices is shown in Figure \ref{fig:lnepsResum}. Since we work at fixed $R=0.2$, the $\ln(R)$ resummation amounts to an overall reduction of the cross section. The $\ln(\epsilon_\gamma )$ become important for $\epsilon_\gamma \lesssim 0.1$. The plot shows fixed-order results for smooth-cone isolation with $n=1$, but our leading-logarithmic (LL) resummation of NGLs is insensitive to the isolation prescription since it only depends on the isolation via the associated scale $\mu_0$. The 
isolation requirement changes the one-loop term \eqref{eq:Uq1}, but this is a NLL effect. 

{\sc ATLAS} imposes $E_0=\epsilon \, E^T_\gamma+ E^T_{\rm th}$ with $E^T_{\rm th}=4.8\, {\rm GeV}$ and $\epsilon = 0.0042$, which which corresponds to a value $\epsilon_\gamma =E_0/E^T_\gamma  \approx 0.04$ for $E^T_\gamma= 125\,{\rm GeV}$. The threshold term $E^T_{\rm th}$ is added by {\sc ATLAS} to avoid that $E_0$ reaches non-perturbative values, but our analysis makes it clear that the lowest scale in the problem is $R\,E_0$ which is close to $1\,{\rm GeV}$ for $R=0.2$ and $E^T_\gamma= 125\,{\rm GeV}$. This  corresponds to a value $t \approx 0.066$. (The value of $t$ very slowly increases for larger $E^T_\gamma$ and reaches $t \approx 0.07$ for $E^T_\gamma=1\,{\rm TeV}$.) Figure \ref{fig:lnepsResum} shows that for $R=0.2$ the resummation lowers the cross section by about $39\,{\rm pb}$, about half of which is due to $\ln(R)$ resummation, while the other half is due to $\ln(\epsilon_\gamma)$ terms. For $R=0.4$, the $\ln(R)$ resummation effects are about half as large, while the size $\ln(\epsilon_\gamma)$ remains about the same.

\section{Summary and conclusion\label{sec:conclusion}}

In this paper, we have studied in detail the structure of QCD effects associated with  isolation requirements imposed in experimental measurements of photon production at high-energy colliders. We have have shown that for small cone radius $R$, 
 the isolation effects can be described by cone fragmentation functions $\mathcal{F}_{i\to \gamma}$ describing the transition of an energetic quark or gluon into a photon plus accompanying QCD radiation.  For small isolation energy $E_0 = \epsilon_\gamma E_\gamma$, these fragmentation function factorize further into jet functions $\bm{\mathcal{J}}_{\!\! i \to \gamma+l}$ describing the $l$ energetic partons outside the isolation cone boundary, and Wilson line matrix elements $\bm{\mathcal{U}}_{l}$ encoding the soft radiation emitted from these partons into the cone. Our factorization theorem separates the different scales present in the cross section: the hard scale $\mu_h \sim E_\gamma$, the jet scale $\mu_j \sim R E_\gamma$ and the isolation energy scale $\mu_0 \sim R E_0$. Using RG methods, we have resummed the leading logarithms of $R$ and the non-global logarithms of $\epsilon_\gamma$. The renormalization group also lets us evaluate each contribution at its natural scale. 
 
 To avoid low scales in the relevant perturbative computations current experimental measurements impose $E_0\gtrsim 5\, {\rm GeV}$, but our analysis demonstrates that for low jet radii one still reaches the dangerously low scale $\mu_0 \sim R E_0$. Values around $R= 0.2$ are commonly used in diphoton measurements, for example in \cite{CMS:2014mvm,ATLAS:2021mbt}. The presence of the low scale $\mu_0$ is problematic for precision computations of photon production and higher-order problems might not immediately be visible since the  isolation is a NLO effect which only affects a certain region of phase space. Even if the perturbative expansion fails for the isolation effects this might not yet be visible at NNLO, since there are other higher-order effects which are of the same size. Indeed, the NNLO cross sections are higher than the NLO results while the resummation effects we computed reduce the cross section. 
 
 Another simple but important result of our analysis is that the effect of non-perturbative fragmentation is suppressed by $\epsilon_\gamma$. For small isolation energy this effect can thus be neglected, which is good news since the non-perturbative fragmentation functions are quite poorly known. 
 
 Our formalism cannot only be used to perform resummation, but also to convert results from one isolation prescription to another.  Indeed, in our paper we have often discussed the difference between cross sections since it is directly proportional to the cone fragmentation functions. An interesting application of our fragmentation framework is to convert NNLO results computed with smooth-cone isolation to results in the fixed-cone scheme. We have presented a simple formula, which achieves this conversion in the limit of small $\epsilon_\gamma$, which should be sufficient for most applications. Interestingly, the cross section difference is proportional to $n \,\zeta_3$, where $n$ is the parameter of smooth-cone isolation. One could extend this result to arbitrary $\epsilon_\gamma$ by computing $\mathcal{F}_{i\to \gamma}$ at NLO. 
 
Our computations were carried out in RG-improved perturbation theory at NLO. Since the fragmentation contribution only arises at $\mathcal{O}(\alpha_s)$, this corresponds to NLL resummation of the $\ln(R)$ terms and LL resummation of the $\ln(\epsilon_\gamma)$ contributions. To match the accuracy of fixed-order NNLO computations, we should extend the resummation to  subleading logarithms of $R$ and $\epsilon_\gamma$. For the $\ln(R)$ resummation, the $\alpha_s^2$ corrections to $P_{q \to \gamma}(z)$ and $P_{g \to \gamma}(z)$ are as of yet unknown and would need to be computed. For the resummation of $\ln(\epsilon_\gamma)$ the most important ingredient, namely the two-loop evolution, is available \cite{Banfi:2021owj,Banfi:2021xzn,Becher:2021urs}. Only the one-loop boundary conditions, in particular the jet function with an additional parton, will need to be determined and implemented. We look forward to doing so in the future.

\begin{acknowledgments}     
The authors thank Daniel de Florian, Thomas Gehrmann,  Alexander Huss, Tobias Neumann and Ze Long Liu for interesting discussions and comments. We are grateful to Alexander Huss for providing benchmark cross section numbers for {\sc NNLOjet} and to John Campbell and Tobias Neumann for help with {\sc MCFM}. TB would like to thank the Pauli Center at ETHZ and the CERN theory department for hospitality during the completion of this work. This research is supported by the Swiss National Science Foundation (SNF) under grant 200020\_182038.
\end{acknowledgments}  

\begin{appendix}

\section{Splitting functions}\label{app:splitting}

The expansion coefficients of the splitting functions were defined in \eqref{eq:splitex1} and  \eqref{eq:splitex2}. The well-known leading-order QCD splitting functions are 
\begin{align}
P^{(1)}_{q \to q}(z)&=P^{(1)}_{\bar{q} \to \bar{q}}(z)=C_F\left( \left(1+z^2\right) \left [\frac{1}{1-z} \right]_{+}+\frac{3}{2} \delta(1-z) \right) \, , \nonumber
\\
P^{(1)}_{g \to g}(z)&=C_A\left( z \left [\frac{1}{1-z} \right]_{+} +\frac{1-z}{z}+z(1-z) \right)+\frac{\beta_0}{2} \delta(1-z) \, ,\nonumber
\\
P^{(1)}_{g\to q}(z)&=P^{(1)}_{g\to\bar{q}}(z)=T_F\left(z^2+(1-z)^2\right) \, , \nonumber
\\
P^{(1)}_{q \to g}(z)&=P^{(1)}_{\bar{q} \to g}(z)=C_F  \, P(z) = C_F \frac{1+(1-z)^2}{z}\, ,
\end{align}
with $\beta_0 = \frac{11}{3}N_c -  \frac{4}{3} n_f T_F$. The coefficients of the parton-to-photon splitting kernels can be found in \cite{Gehrmann-DeRidder:1997fom} and are given by 
\begin{equation}
\begin{aligned}
P^{(0)}_{g \to \gamma}(z)&=0 \, , 
\\
P^{(0)}_{q \to \gamma}(z)&=P^{(0)}_{\bar{q} \to \gamma}(z)=Q_q^2\,  P(z) \, ,  
\\
P^{(1)}_{q \to \gamma}(z)&=P^{(1)}_{\bar{q} \to \gamma}(z)=\frac{C_F Q_q^2}{2}  \left( - \frac{1}{2} + \frac{9}{2} z + \left( \frac{z}{2} -8 \right) \ln z + 2 z \ln (1-z) + \left( 1- \frac{z}{2} \right) \ln^2 z   \right. \\
&\left. + \left[ \ln^2(1-z)  + 4 \ln z \ln (1-z) + 8\,{\rm Li}_2 (1-z) - \frac{4 \pi^2 }{3}\right] P(z)  \right) \\
P^{(1)}_{g \to \gamma}(z)&= \frac{T_F \sum_{q=1}^{n_f} Q_q^2 }{2} \left( -2 + 6 z -\frac{82}{9} z^2  + \frac{46}{9 z } + \left( 5 + 7 z+ \frac{8}{3} z^2+ \frac{8}{3 z}  \right) \ln z  + (1+z) \ln^2 z \right) \, .  
\end{aligned}
\end{equation}
The factor $ \sum_{q=1}^{n_f} Q_q^2$ is due to a quark loop and is $11/9 $ for  $n_f=5$ quark flavors. We note that the kernels $P^{(1)}_{q \to \gamma}(z)$ and $P^{(1)}_{g \to \gamma}(z)$ differ from the ones relevant for the space-like case, which are given in \cite{deFlorian:2015ujt}. Note that compared to the expressions in \cite{Gehrmann-DeRidder:1997fom,deFlorian:2015ujt} we have an additional factor $\frac{1}{2}$ in $P^{(1)}_{q \to \gamma}$ and $P^{(1)}_{g \to \gamma}$ due different conventions: these references expand in $\frac{\alpha_s}{2\pi}$ instead of $\frac{\alpha_s}{\pi}$ and write the evolution equations in the variable $\mu^2$ instead of $\mu$.

\section{Solution of the RG equations of the fragmentation functions}
\label{app:momentSol}
In order to solve \eqref{eq:DGLAPmom}, we perform a transformation of the fragmentation function such that the differential equations are decoupled from each other, i.e.\ one performs a basis change
\begin{equation}
\label{eq:decouplingN}
\hat{\mathcal{F}}_{i\to \gamma}(N,\mu) = U_{ij}(N) \mathcal{F}_{j\to \gamma}(N,\mu)
\end{equation}
and chooses the matrix $U_{ij}(N)$ in such a way that the splitting kernel becomes diagonal
\begin{equation}
U_{ij}(N) \mathcal{P}_{j\to k}(N, \mu)  U_{kl}^{-1}(N) = \hat{{\mathcal P}}_{i\to i}(N, \mu) \delta_{il}\,.
\end{equation}
This diagonalization step is only necessary for $\Sigma$ and $G$ in \eqref{eq:Sigma} since $\Delta$ is already decoupled from the other two quantities. The diagonalized evolution equation \eqref{eq:DGLAPmom} takes the form
\begin{align}
\frac{d}{d\ln\mu} \hat{\mathcal{F}}_{i\to \gamma}(N,\mu) &= \hat{ {\mathcal P}}_{i\to \gamma}(N, \mu) +  \hat{{\mathcal P}}_{i\to i}(N, \mu)  \hat{\mathcal{F}}_{i\to \gamma}(N,\mu) \, .
\end{align}
The solution of this differential equation is
\begin{align}
\hat{\mathcal{F}}_{i\to \gamma}(N,\mu) &= \exp \left[ \int_{\mu_0}^{\mu} d \ln\mu^\prime\, \hat{{\mathcal P}}_{i\to i}(N, \mu^\prime) \right]  \, \nonumber
\\
&\left\{   \hat{\mathcal{F}}_{i\to \gamma}(N,\mu_0)+ \int_{\mu_0}^{\mu} d \ln \mu^{\prime\prime}  \exp\!\left[- \int_{\mu_0}^{\mu^{\prime \prime} } d \ln \mu^{\prime} \, \hat{{\mathcal P}}_{i\to i}(N, \mu^{\prime}) \right] \hat{ {\mathcal P}}_{i\to \gamma}(N,\mu^{\prime\prime}) \right \} \, .
\end{align}
Because $\hat{{\mathcal P}}_{i\to i}(N, \mu)$ only depends on $\mu$ via the strong coupling,
\begin{align}
\hat{{\mathcal P}}_{i\to i}(N, \mu)= \frac{\alpha_s(\mu)}{\pi} \hat{ P}^{(1)}_{i\to i}(N) + \bigg(\frac{\alpha_s(\mu)}{\pi}\bigg)^2 \hat{ P}^{(2)}_{i\to i}(N) +\dots \, ,
\end{align}
we can rewrite the exponential as
\begin{align}\label{eq:kappa}
\mathcal{K}(\alpha_s(\mu_0),&\alpha_s(\mu))=\exp \left[ \int_{\mu_0}^{\mu} d \ln \mu^\prime \hat{{\mathcal P}}_{i\to i}(N, \mu^\prime) \right]
= \exp \left[ \int_{\alpha_s(\mu_0)}^{\alpha_s(\mu)} \frac{d  \alpha}{\beta(\alpha)} \hat{{\mathcal P}}_{i\to i}(N, \mu^\prime) \right] \, \nonumber
\\
& = \left(  \frac{\alpha_s(\mu_0)}{\alpha_s(\mu)} \right)^{\frac{2 \hat{ P}^{(1)}_{i\to i}(N)}{\beta_0} } \left[ 1+ \frac{\alpha_s(\mu)-\alpha_s(\mu_0)}{4 \pi} \frac{2}{\beta_0} \left(\frac{\beta_1}{\beta_0}\hat{ P}^{(1)}_{i\to i}(N) -4 \hat{ P}^{(2)}_{i\to i}(N)\right) \right]
\end{align}
in which we have expanded
\begin{equation}
\beta(\alpha)=-2\alpha_s  \left(\beta_0\, \frac{\alpha_s}{4 \pi}+ \beta_1 \left(\frac{\alpha_s}{4 \pi}\right)^2 +\dots \right)
\end{equation}
and dropped higher-order terms. The solution can thus be rewritten as 
\begin{align}
\hat{\mathcal{F}}_{i\to \gamma}(N,\mu) &=\mathcal{K}(\alpha_s(\mu_0),\alpha_s(\mu))  \left[ \hat{\mathcal{F}}_{i\to \gamma}(N,\mu_0)+     \int_{\alpha_s(\mu_0)}^{\alpha_s(\mu)} \frac{d  \alpha}{\beta(\alpha)} \mathcal{K}(\alpha,\alpha_s(\mu_0)) \hat{ {\mathcal P}}_{i\to \gamma}(\alpha) \right] \, ,
\end{align}
where we suppressed the argument $N$ of the splitting function on the right-hand side of the equation. 
Expanding also the inhomogeneous $\hat{ {\mathcal P}}_{i\to \gamma}(N,\mu)$ as
\begin{align}
\hat{ {\mathcal P}}_{i\to \gamma}(N,\mu)=\frac{\alpha_{\text{EM}}}{\pi}\left( \hat{P}^{(0)}_{i\to \gamma}(N)+\frac{\alpha_s}{\pi} \hat{P}^{(1)}_{i\to \gamma}(N) \right) \, ,
\end{align}
the final form of the solution reads 
\begin{align}\label{eq:solution}
\hat{\mathcal{F}}_{i\to \gamma}(N,\mu) &= \frac{2 \,\alpha_{\text{EM}} \,\hat{P}^{(0)}_{i\to \gamma}}{\alpha_s(\mu)(\beta_0-2\hat{P}^{(1)}_{i\to i})} \left(1-\frac{\alpha_s(\mu)}{\alpha_s(\mu_0)} \mathcal{K}(\alpha_s(\mu_0),\alpha_s(\mu))\right) \, \nonumber
\\
&-\frac{\alpha_{\text{EM}} }{\pi }\left(\frac{\hat{P}^{(1)}_{i\to \gamma}}{\hat{P}^{(1)}_{i\to i}} -\frac{\beta_1}{4\beta_0}\frac{\hat{P}^{(0)}_{i\to \gamma}}{\hat{P}^{(1)}_{i\to i}}  -\frac{\beta_1\hat{P}^{(0)}_{i\to \gamma}}{2\beta_0\left(\beta_0-2\hat{P}^{(1)}_{i\to i} \right)}\right) \left(1-\mathcal{K}(\alpha_s(\mu_0),\alpha_s(\mu)) \right) \, \nonumber
\\
&+\mathcal{K}(\alpha_s(\mu_0),\alpha_s(\mu))\hat{\mathcal{F}}_{i\to \gamma}(N,\mu_0) \, ,
\end{align}
The first term in this solution corresponds to LL resummation and is proportional $1/\alpha_s$, the remaining two terms are the NLL corrections. In these two terms one can omit the $\mathcal{O}(\alpha_s)$ corrections to $\mathcal{K}$ in \eqref{eq:kappa}. In our numerical evaluation, we do not include $\beta_1$ terms and the corrections from $\hat{ P}^{(2)}_{i\to i}(N)$ in the first line of \eqref{eq:solution} for simplicity, even though they would formally be needed for NLL accuracy and are available \cite{Gluck:1992zx}. We verified that the $\beta_1$ terms are numerically very small.

After solving for $\hat{\mathcal{F}}_{i\to \gamma}(N,\mu) $ we first undo the decoupling change of variables \eqref{eq:decouplingN} and go back to the original functions $\mathcal{F}_{j\to \gamma}(z,\mu)  $. We then use the same Mellin inversion contour as in \cite{Graudenz:1995sk} to transform the $\hat{\mathcal{F}}_{i\to \gamma}(N,\mu)$ back to $z$ space: 
\begin{equation}  \label{eq:inverseMellin}
   \mathcal{F}_{j\to \gamma}(z,\mu) = \frac{1}{\pi} \int_0^{\infty} \mathrm{d} r  \, \mathrm{Im} \left[ e^{i \phi} z^{-c-r e^{i\phi }} \hat{\mathcal{F}}_{j\to \gamma} \left( N = c + r e^{i \phi} ,\mu \right) \right],
\end{equation}
where $\phi$ and $c$ are parameters chosen as $\phi=\frac{3\pi}{4}$ and $c=1.8$ in our numerical evolution code.

\section{Computation of $\Delta \sigma$\label{app:dSigma}}

To study the dependence on the parameters $ \left( \epsilon_{\gamma}, n , R \right)$ of the smooth-cone isolation \eqref{eq:smooth} we compute the difference to a reference cross section,
\begin{equation*}
    \Delta \sigma = \sigma \left( \epsilon_{\gamma}, n , R  \right) - \sigma  ( \epsilon_{\gamma}^{\re},n^{\re}, R^{\re} )\,.
   \end{equation*}
In order for the difference to be be positive we require that the reference isolation $\left( \epsilon_{\gamma}^{\re},n^{\re}, R^{\re}  \right) $ is the more restrictive one and we impose
\begin{equation}
    \epsilon_{\gamma}  \geq \epsilon_{\gamma}^{\re} \quad, \quad n \leq n^{\re} \quad, \quad R \leq R^{\re} \,.
\end{equation}
In this appendix, we provide details on the fixed-order determination of $ \Delta \sigma$ and the computation of the difference based on cone fragmentation functions using \eqref{eq:diffFrag}. 

\subsection{Event-based fixed-order computation}

Rather than computing the difference of NLO photon production cross sections, it is much more efficient to directly extract the cross section difference from the process $p \, p \rightarrow \gamma\, j\, j $ at LO by imposing suitable cuts on the partons. Working in this way, we can compute the difference from event files generated with {\sc MadGraph5\_aMC@NLO}  \cite{Alwall:2014hca} instead of needing individual NLO runs for all parameter values. 

\begin{figure}
    \centering
\includegraphics[width=6cm]{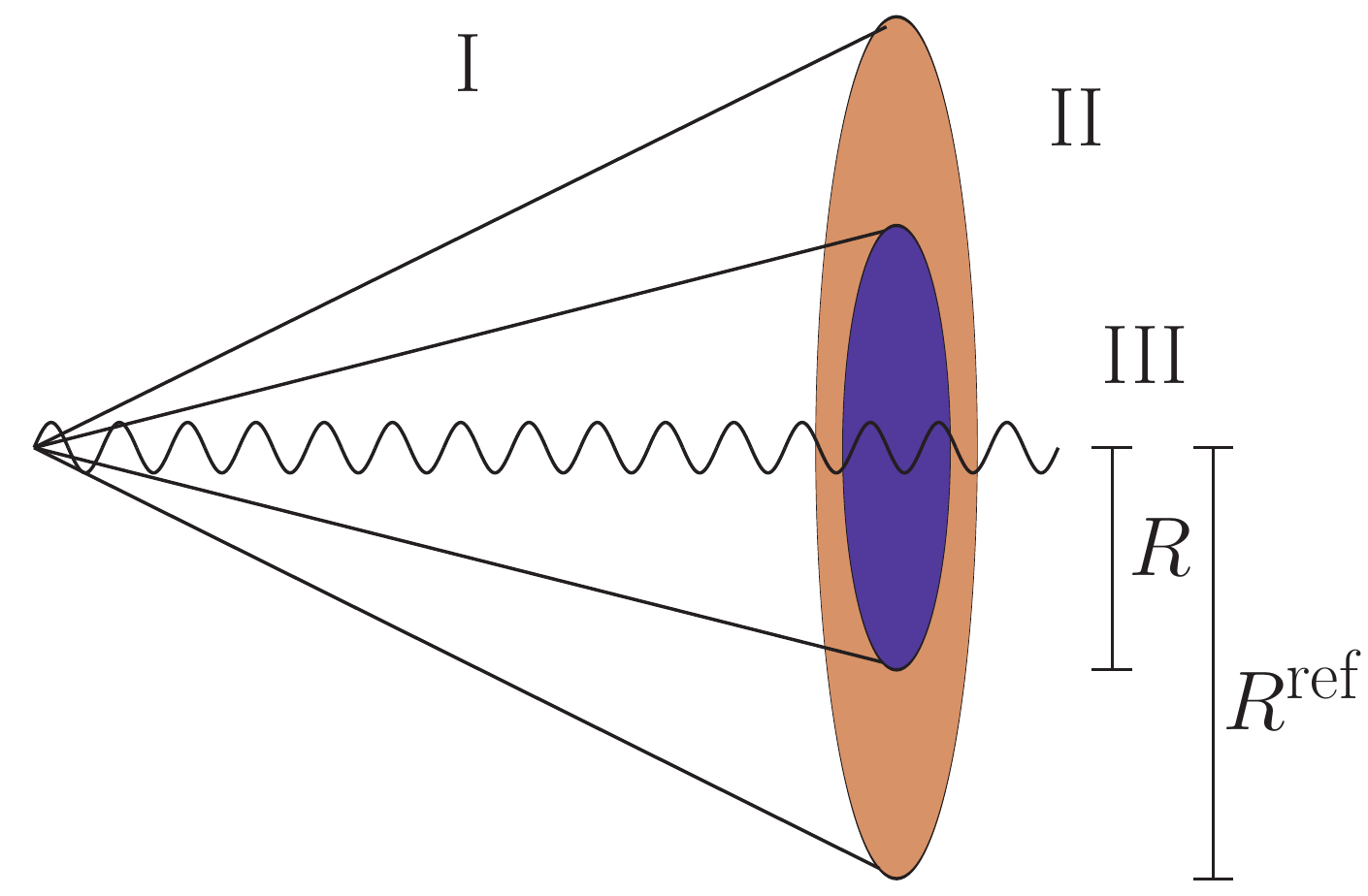}
\caption{Isolation cones of radius $R$ and $R^{\re}$ and associated angular regions. The difference $\Delta \sigma$ is obtained  from partonic configurations which fulfill the isolation criterion for cone radius $R$, but fail it for  $R^{\re}$. \label{fig:cones}}
\end{figure}

In order to get a contribution to $ \Delta \sigma$ we need that at least one of the partons in the $p \, p \rightarrow \gamma\, j\, j $ event to be inside the larger cone with radius $R^{\re}$. The second parton will be outside the cone, since it is recoiling against the energetic photon. For an event to contribute to $ \Delta \sigma$ it should respect the constraint imposed by isolation $ \left( \epsilon_{\gamma}, n , R \right)$ but fail the one with the reference values $ \left( \epsilon_{\gamma}^{\re},n^{\re}, R^{\re}  \right) $. To formulate the resulting constraints on the transverse momenta $p_i^T$ of the two  final-state QCD partons $i\in \{1,2\} $ in the event explicitly, we distinguish three angular regions indicated in Figure \ref{fig:cones}:
 \begin{enumerate}[label=(\Roman*)]
\item$r_i>R^{ \re}$: at most one parton, no constraint on $p_i^T$,
 \item $ R<r_i<R^{\re}$: $ \epsilon_{\gamma}^{\re} \left(  \frac{1- \cos r_i}{1-\cos R^{\re} }\right)^{n^{\re}}  \leq \frac{p_i^T}{p_{\gamma}^T}$ ,
\item  $r_i<R$:     $ \epsilon_{\gamma}^{\re} \left(  \frac{1- \cos r_i}{1-\cos R^{\re} }\right)^{n^{\re}}  \leq \frac{p_i^T}{p_{\gamma}^T} \leq   \epsilon_{\gamma} \left(  \frac{1- \cos r_i}{1-\cos R }\right)^{n}   $, 
\end{enumerate}
where $r_i$ is the angular distance of the parton to the photon. The implementation of these non-standard cuts in {\sc MadGraph5\_aMC@NLO} is achieved by modifying the file {\tt cuts.f}. 
\subsection{Fragmentation contribution}

The fixed-order results are then compared with \eqref{eq:diffFrag} based on the factorization theorem \eqref{factorizationFormulaSmallCone}. According to \eqref{eq:diffFrag} the leading contribution to the difference of cross sections is given by the partonic cross section $d\sigma_{i+X}/dE_i$ convoluted with the difference of fragmentation functions 
\begin{equation}
\Delta \mathcal{F}_{i \rightarrow \gamma} =  \mathcal{F}_{i \rightarrow \gamma}  \left( z,R,\epsilon_{\gamma},n\right)  -  \mathcal{F}_{i \rightarrow \gamma} \left(z,R^{\re} ,\epsilon_{\gamma}^{\re},n^{\re} \right) \, .
\end{equation}
We obtain the partonic cross section by using  {\sc MadGraph5\_aMC@NLO} to generate event files for the process $ p\, p \to j \,j$ at leading order. These have two QCD partons in the final state, each one of which can then fragment into a photon.

In the main text, we computed and plotted $\Delta \sigma$ for three different cases and we now list the relevant $\Delta \mathcal{F}_{i \rightarrow \gamma}$.  To study the $n$-dependence of $\Delta \sigma$, we set $R=R^{\re}$ and $\epsilon_{\gamma}=\epsilon_{\gamma}^{\re}$ which yields
\begin{equation}
\begin{aligned}
 \Delta \mathcal{F}_{i \rightarrow \gamma} &=   \frac{ \alpha_{\rm EM} Q_i^2 }{2 \pi }P(z) \, \theta\!\left(z-\frac{1}{1+\epsilon_{\gamma} } \right)  \ln \left( \frac{1-z}{z \epsilon_{\gamma}}\right)\left( \frac{1}{n}  - \frac{1}{n^{\re} }   \right)\,.
\end{aligned}
\end{equation}
To study the  $R$ dependence, we set $\epsilon_{\gamma} = \epsilon_{\gamma}^{\re}$ and $n=n^{\re}$ which leads to
\begin{equation}
\Delta \mathcal{F}_{i \rightarrow \gamma}  = \frac{ \alpha_{\rm EM} Q_i^2 }{ \pi }  P(z)   \ln \left( \frac{R^{\re}}{R}  \right)\,.
\label{eq:DeltaF_Frixione_Rdep}
\end{equation}
The most complicated case is the $\epsilon_{\gamma}$-dependence for which the relevant $\Delta \mathcal{F}_{i \rightarrow \gamma}$ for  $R=R^{\re}$ and $n=n^{\re}$ was given in the main text in \eqref{eq:deltaFepsgamma}.

\section{Reference cross section values}\label{app:ref}

In the main text we used {\sc MadGraph5\_aMC@NLO}  \cite{Alwall:2014hca} and {\sc MCFM} \cite{Campbell:2019dru} for our computations. The authors of \cite{Chen:2022gpk}  have compared their {\sc NNLOjet}  fixed-energy cone results at NLO to the {\sc JetPhoX} code \cite{Catani:2002ny} and have provided reference cross section numbers for fixed-cone isolation in their paper. We have  verified that {\sc MCFM} reproduces the reference cross section in \cite{Chen:2022gpk} within numerical uncertainties. The authors of \cite{Chen:2022gpk} were kind enough to also provide us with reference cross section numbers for smooth-cone isolation with $R=0.4$, $n=1.0$ and $\epsilon_{\gamma}=0.0042$ and we have verified that all the above codes produce compatible results within numerical uncertainties. These reference cross sections were computed for  $\alpha_{\mathrm{EM}} = 1/137$ and NNPDF31\_nnlo\_as\_0118\_mc PDFs, after imposing $E_{\gamma}^T \geq 125\,{\rm GeV}$, $|\eta_{\gamma}| \leq 2.37$ and setting $\mu_F=\mu_R=p_{\gamma}^T$. For the leading order cross section and the NLO correction, they obtain
\begin{align}
\sigma^{\rm LO} &= (192.524\pm 0.015) \,{\rm pb} \,, & \Delta \sigma^{\rm NLO} &= (163.44  \pm  0.11) \,{\rm pb}\,.
\end{align}
After requiring at least one jet with $p_{\mathrm{jet}}^T \geq 100\,{\rm GeV}$ and $ |\eta_{\mathrm{jet}}| \leq 2.37$ defined using the $k_T$ algorithm with $R_{\mathrm{jet}}=0.4$, the NLO correction reduces to
\begin{align}
 \Delta \sigma^{\rm NLO} &= (121.441 \pm 0.065) \,{\rm pb}\,.
\end{align}

\end{appendix}

\end{document}